\documentclass[12pt]{article}

\usepackage{epsfig}

\def\be{\begin{equation}}
\def\ee{\end{equation}}
\def\bea{\begin{eqnarray}}
\def\eea{\end{eqnarray}}
\def\ii{{\rm i}}
\def\tr{{\rm Tr}}
\def\nn{\nonumber}
\def\ex{{\rm e}}
\def\lsi{\raise0.3ex\hbox{$<$\kern-0.75em\raise-1.1ex\hbox{$\sim$}}}
\def\gsi{\raise0.3ex\hbox{$>$\kern-0.75em\raise-1.1ex\hbox{$\sim$}}}

\newcommand{\gsim}{\mathop{\gsi}}
\newcommand{\eq}{Eq.~}

\newcommand{\op}{{\cal O}}
\newcommand{\bmu}{\bar{\mu}}
\def\bfx{{\bf x}}

\title{
\bf Status of Lattice QCD at \\ Finite Temperature}
\author{Edwin Laermann$^a$ and Owe Philipsen$^b$ \\[5mm]
        $^a$ \small Fakult\"at f\"ur Physik, Universit\"at Bielefeld, \\
        \small D-33615 Bielefeld, Germany \\[2mm]
        $^b$ \small Center for Theoretical Physics, 
        Massachussetts Institute of Technology, \\
        \small Cambridge MA 02139, USA
       \vspace*{0.5cm}}
\date{\small  BI-TP 2003/06, MIT-CTP-3345}

\begin{document}

\maketitle


\begin{abstract}
\noindent
The status of lattice QCD investigations at 
high temperature is reviewed.
After a short introduction into thermal QCD
on the lattice we report on the present understanding
of the phase diagram and the equation of state,
in particular in presence of dynamical quarks.
We continue with a discussion of various screening
lengths in the plasma phase including results
from dimensionally reduced QCD. This is followed
by summarizing lattice data on quark number susceptibilities
and spectral densities, both of which are of immediate relevance
to the interpretation of heavy ion experiments.
A major section is devoted to presenting simulations
of QCD at small yet phenomenologically important
values for the baryon density.

\end{abstract}

\newpage
\tableofcontents

\newpage
\section{Introduction}

Quantumchromodynamics (QCD) is the theory describing the strong interactions, carried
by gluons, which confine quarks into hadrons.
It is the fundamental theory underlying all nuclear physics, 
and as such responsible for many salient features of matter
as we know it today. 
A key property of the theory is asymptotic freedom, according to which
the coupling strength decreases with the energy transfer of an interaction.
One consequence is that
at high enough energies 
of a few GeV                                      
the theory is tractable by perturbation
theory, which leads to its experimental verification in 
e.g. deep inelastic scattering.
On the other hand, on the low energy scales of hadron physics the coupling is strong
and perturbation theory invalid. The only known non-perturbative and
first principles method to compute QCD predictions in this regime is by simulations of
lattice gauge theory, whose results are beginning to give a quantitative description
of the hadron spectrum.

At a critical temperature $T_c\sim 200$ MeV, 
QCD predicts a phase transition to a deconfined plasma 
of quarks and gluons. At the same temperature the weakly broken chiral symmetry, responsible
for the existence of light pions, gets restored. Such temperatures were
realized in the early universe, which during its cooling expansion passed through the plasma
phase and the quark hadron transition on its way to its present state. 
We are currently witnessing
exciting attempts to recreate this primordial plasma and its subsequent cooling through
the phase transition in heavy ion collision experiments at 
RHIC (BNL) as well as SPS and LHC (CERN).
These studies will have a bearing far beyond QCD in the context of early universe physics.
Many prominent features of the observable universe, such as the baryon asymmetry or the seeding
for structure formation, have been determined primordially 
in hot plasmas described by non-abelian gauge theories. The QCD plasma is prototypical 
for those as it is the only one we can hope to produce in laboratory experiments.
Moreover, for high densities and low temperatures more exotic non-hadronic phases
like a color superconductor have been predicted, and 
there is a chance for such physics to be realized in the
cores of compact stars.

For such applications we need to know how the
properties of QCD change under extreme conditions of temperature and density.
This entails a determination of the QCD phase diagram and the associated critical 
properties, a quantitative understanding of the equation of state, the way
in which hadron properties get modified as well as the nature of the lightest excitations
in non-hadronic phases. Answers to those questions may also help
to elucidate mechanisms of confinement and chiral symmetry breaking.

In this review we collect data from lattice simulations 
of thermal QCD and try to place them into their physics context. 
While avoiding technical issues as far as possible,
our aim is to convey
a picture of what we have learned about the QCD transition and plasma, even though
in places this picture is still incomplete. 
When the critical temperature is crossed, light degrees
of freedom appear to be released, in accordance with the naive expectation of deconfinement. 
However, the resulting plasma is not compatible with the perturbative picture
of a free parton gas for any temperatures of interest. 
Rather, the system displays strong residual intercations 
caused by soft modes in the gauge sector. 

Static equilibrium physics is 
well under control in the continuum limit for the pure gauge theory without
quarks, whereas calculations with dynamical quarks are limited to 
$m_{\pi}\gsim 300$ MeV, and even those are still affected by cut-off
effects. Nevertheless, these systematic errors can be reduced step by step in
additional calculations to come.

Less advanced and not yet fully quantitative are 
the calculations in two other areas of interest.
Since lattice QCD works in euclidean space-time, it is much harder to obtain 
a real time description of dynamical processes. 
This field is in its infancy and we
will report on a first                                
step taken in that direction.
In the last two years significant progress has been made towards
simulating the regime of small baryon densities relevant for heavy ion 
collisions, which was entirely impossible before then. While this also 
is a rather young field of research with a long way to go,
there now are first results for the $(\mu,T)$ phase diagram and the search
for a potential critical endpoint of the deconfinement transition.
Before presenting lattice results available today, 
we give a brief introduction to the theoretical set-up of lattice thermodynamics
and a discussion of its current limitations.


\section{QCD Thermodynamics on the Lattice}
\label{sec:thermo_lat}

Detailed introductions to thermal field theory and lattice field theory
can be found in \cite{bellac} and \cite{munster}, respectively.
The QCD grand canonical partition function is given by
\be
Z(V,\mu,T)=\tr \left(\ex^{-(\hat{H}-\mu \hat{Q})/T}\right)=
\int DA D\bar{\psi}D\psi \, \ex^{-S_g[A_\mu]} \ex^{-S_f[\bar{\psi},\psi,A_\mu]},
\ee
with the euclidean gauge and fermion actions
\bea
S_g[A_\mu]&=& \int\limits_0^{1/T} dx_0 \int\limits_V d^3 {\bf x} \;
\frac{1}{2} {\rm Tr}\; F_{\mu\nu} F_{\mu\nu}, \nn \\
S_f[\bar{\psi},\psi,A_\mu]&=& \int\limits_0^{1/T} dx_0 \int\limits_V d^3 {\bf x} \;
 \sum_{f=1}^{N_f} \bar{\psi}_f \left( \gamma_\mu
D_\mu+ m_q^f- \mu \gamma_0 \right) \psi_f .
\label{lagrangian}
\eea
Its thermodynamic parameters are the temperature $T$, the volume $V$ and 
the chemical potential $\mu$
for quark number $Q$ (the one for baryon number is $\mu_B=3\mu$),
while the QCD action depends on the number of quark flavors $N_f$, their masses $m^f_q$
and the gauge coupling $g$. 
In the following we will consider mostly two and three flavors,
and always take $m_u=m_d$. The case $m_s=m_{u,d}$ is then denoted by $N_f=3$, while $N_f=2+1$
implies $m_s\neq m_{u,d}$.

\subsection{The Lattice Formulation}

The theory is discretized by introducing a euclidean space-time 
lattice $L^3\times N_t$ with lattice spacing $a$, such that volume and temperature are
\be \label{latvt}
V=(aL)^3,\qquad T=\frac{1}{aN_t}\,.
\ee
The fermion fields live on the lattice sites $x$, whereas the gauge fields are represented by
link variables $U_\mu(x)\in SU(3)$ connecting the sites. After a suitable discretization of
the actions, the Gaussian integral over the quark fields can be performed leading to the
lattice partition function
\be
Z(L,a\mu,N_t;\beta,N_f,am_q^f)=
        \int DU\, \prod_f \det M(\mu)\ex^{-S_g[U]},
\ee
where $M(\mu)=\gamma_\mu
(D_\mu+ m_f- \mu \gamma_0)$ 
is the fermion matrix
and 
$S_g[U]$ denotes the gauge               
action with lattice gauge coupling 
$\beta=6/g^2$.
For $\mu=0$, the theory in this form is amenable to a stochastic 
calculation of expectation values
by Monte Carlo methods. Simulations at $\mu\neq 0$ have only begun to become
possible in certain limits, as will be discussed in detail in 
Section \ref{sec:mu}.

The lattice spacing depends on the gauge coupling, $a = a(g)$,
and is set by
some physical (zero temperature) quantity like a hadron mass. 
The physical temperature is then obtained via \eq (\ref{latvt}) as
$T/m_H=1/(a m_H N_t)$. 
In general, calculations are affected by finite size and cut-off effects, the latter
depending on the particular discretization chosen.
The task then is to compute observables for various volumes
and lattice spacings, and extrapolate to the thermodynamic ($V\rightarrow \infty$)
and continuum limits ($a\rightarrow 0,N_t\rightarrow \infty$),
while keeping $T$ and physical parameters fixed.

In practice calculations are severely limited by computational resources, 
the most expensive part being the evaluation of the fermion determinant.
For this reason it has often been set to $1$ in the past, corresponding to the quenched
approximation which neglects all quark loop contributions. In dynamical simulations at
finite temperature, the Wilson and staggered fermion actions 
are used predominantly, as they are the
cheapest to simulate. Another choice with better chiral properties are 
domain wall fermions, while dynamical overlap fermions so far have not been employed.

Typical lattice sizes 
of current dynamical simulations
are $16^3\times 8$ or $32^3\times 4$,
but for many exploratory studies smaller lattices have to make do. In order to 
have reasonably small finite size and cut-off effects, 
the correlation length $\xi$ of a typical hadronic state
has to fit comfortably in the box while being much larger than the lattice spacing,
\be \label{lsize}
a\ll \xi\ll aL.
\ee
Around the deconfinement transition $T_c \sim 200$ MeV $\sim (1 {\rm fm})^{-1}$, 
the temporal size $N_t=4-8$ implies lattice spacings of $a\sim 0.1-0.3$ fm, allowing for
box sizes of $aL\sim 1.5-3$ fm.
In the chirally broken phase
the right inequality then constrains the lightest feasible quark masses, implying that
the physical pion cannot be accomodated on such lattices. Typical calculations
correspond to $m_{\pi}\sim 300$ MeV.

Above $T_c$ meson  
correlation lengths typically scale as
$ \xi \sim 1/T$ so that the constraint of \eq (\ref{lsize}) reads
\be \label{xi}
\frac{1}{N_t} \ll 1 \ll \frac{L}{N_t}\;.
\ee
Cut-off effects are then controlled by $1/N_t$
independent of temperature.
These cut-off effects can be rather severe for certain thermodynamic quantities.
The reason is that thermal
distribution functions are peaked for modes with momenta $p\sim T$ of the order of
the cut-off, 
or in lattice units $a p \sim 1/N_t$.
Control over these effects thus warrants the use of improved actions, which are augmented
by additional terms irrelevant in the continuum limit and tuned such that, at finite lattice
spacing, they subtract cut-off effects to a given order $(a^n)$ in the lattice spacing.
In addition, 
cut-off effects can be smaller on
anisotropic lattices, for which a finer spacing
in the temporal direction is chosen, $a_t < a_s$,
corresponding to a larger number of points $N_t$.
However, 
soft gluonic modes scale as $ \xi \sim 1/g^2(T)T$, upon which $1/N_t\rightarrow g^2(T)/N_t$
in \eq (\ref{xi}), 
thus requiring large aspect ratios $L/N_t$ in both cases, which 
limits simulations of very large temperatures.

\subsection{\label{sec:dr}Effective High T Theory: Dimensional Reduction}

At large $T$, when the gauge coupling $g(T$) is sufficiently small, 
a hierarchy between different relevant scales of thermal QCD develops,
\be \label{eq:scale}
2\pi T \gg gT \gg g^2T.
\ee
The lowest non-vanishing bosonic Matsubara mode $\sim 2\pi T$ is characteristic for
non-interacting particles. The dynamics generates the Debye scale $m_E\sim gT$, over which
color electric fields are screened, and its non-perturbative analogue $m_M\sim g^2T$  
for color magnetic fields \cite{linde}.

For physics on scales larger than the inverse temperature, $|\bfx|\sim 1/gT\gg
1/T$, this allows an effective theory approach in which the
calculations are factorized:
integration over the hard modes may be performed perturbatively
by expanding in powers of the ratio of scales $gT/(2\pi T)\sim g/(2\pi)$.
This includes all non-zero Matsubara modes, in particular the fermions.
It results in a 3d effective theory for modes $\sim gT$ and softer, 
\be \label{actc}
        S_{eff} = \int d^{3}x \left\{ \frac{1}{2} \tr(F_{ij}F_{ij})
        +\tr(D_{i}A_0)^2 
+m_E^2 \tr(A_0^2)
        +\lambda_3(\tr(A_0^2)^{2} \right\} ,
\ee 
and is
known as dimensional reduction \cite{dr}. 
Since 4d euclidean time has been integrated over, 
$A_0$ now appears as an adjoint scalar, and the effective
parameters are functions of the original ones, 
$g_3^2=g^2T,m_E(N,N_f,g,m_q^f)\sim gT,\lambda_3(N,N_f,g,m_q^f)\sim g^4T$.

Discretization and simulation of this reduced problem is easy. 
Without fermions and one dimension less,
much larger volumes and finer lattices can be considered.
Moreover, 3d gauge theories are superrenormalizable and the coupling scales linearly with
lattice spacing. Hence, very accurate continuum limits can be obtained and systematic
errors from simulations are practically eliminated.
However, the reduction step
entails two approximations: the perturbative computation is limited to
a finite order in $g$ and neglects higher-dimensional operators, which are
suppressed by powers of the
scale ratio.
The reduction step has been performed up to two-loop order \cite{ad} at which parameters
have relative accuracy ${\cal O}(g^4)$,
while for correlation functions the error is \cite{rules}
$\delta C/C\sim {\cal{O}}(g^3)$.
In the treatment of the electroweak phase transition,
this is less than 5\%\cite{ew},
for QCD it depends on the size of the coupling $g(T)$. As we will discuss in
Section \ref{sec:screen}, it can be
non-perturbatively checked that the effective theory 
accurately describes static correlation lengths
at temperatures as low as 
$T\gsim 2T_c$,
thus allowing for a straightforward treatment of very large temperatures as
well as detailed dynamical investigations in the plasma phase.

\section{The $(m_{u,d},m_s)$ Phase Diagram}
\label{sec:phase}

\begin{figure}[t]
 \begin{minipage}[t]{.450\linewidth}
\epsfig{file=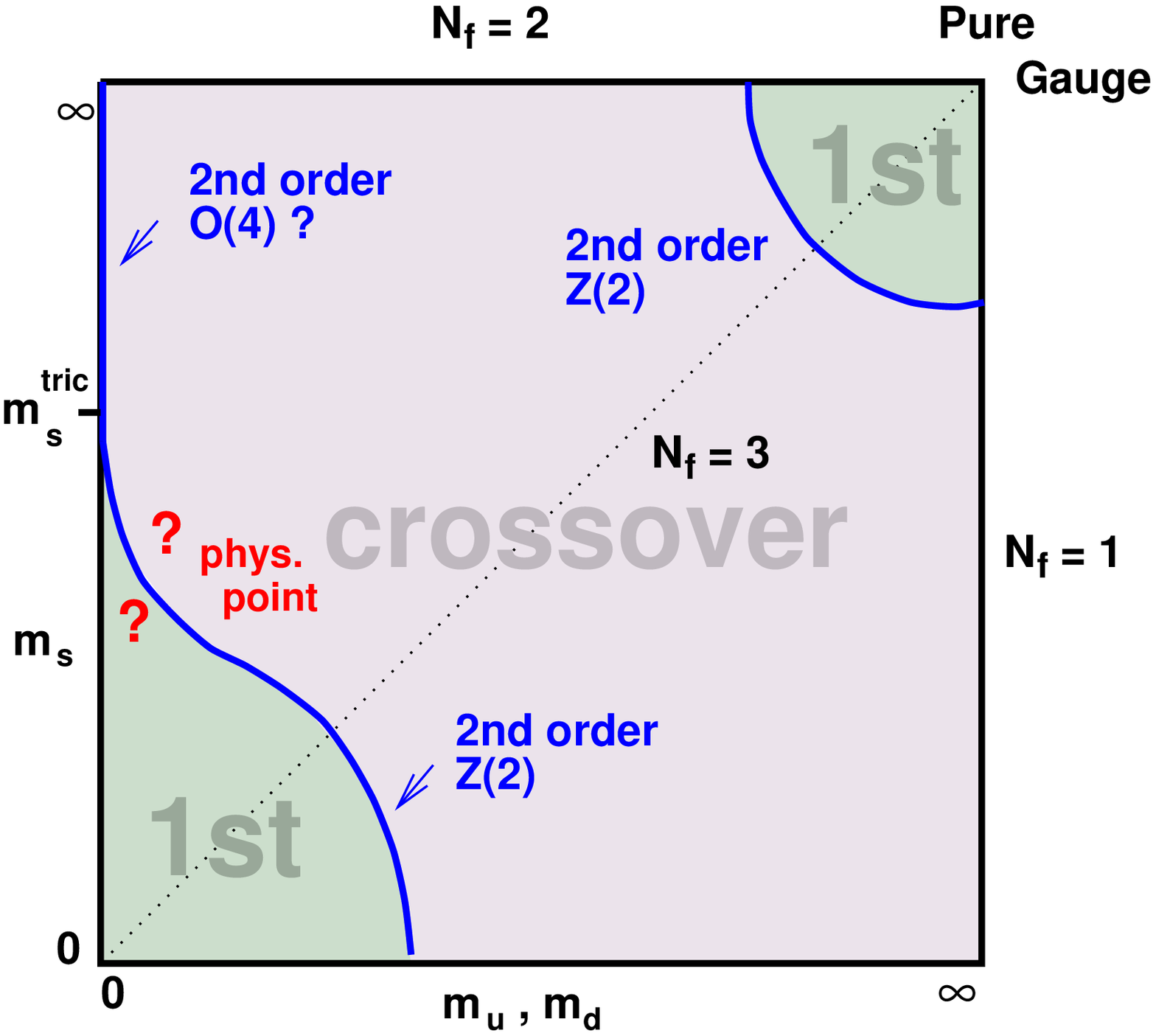,width=65mm}
\end{minipage}
 \begin{minipage}[b]{.500\linewidth}
\epsfig{file=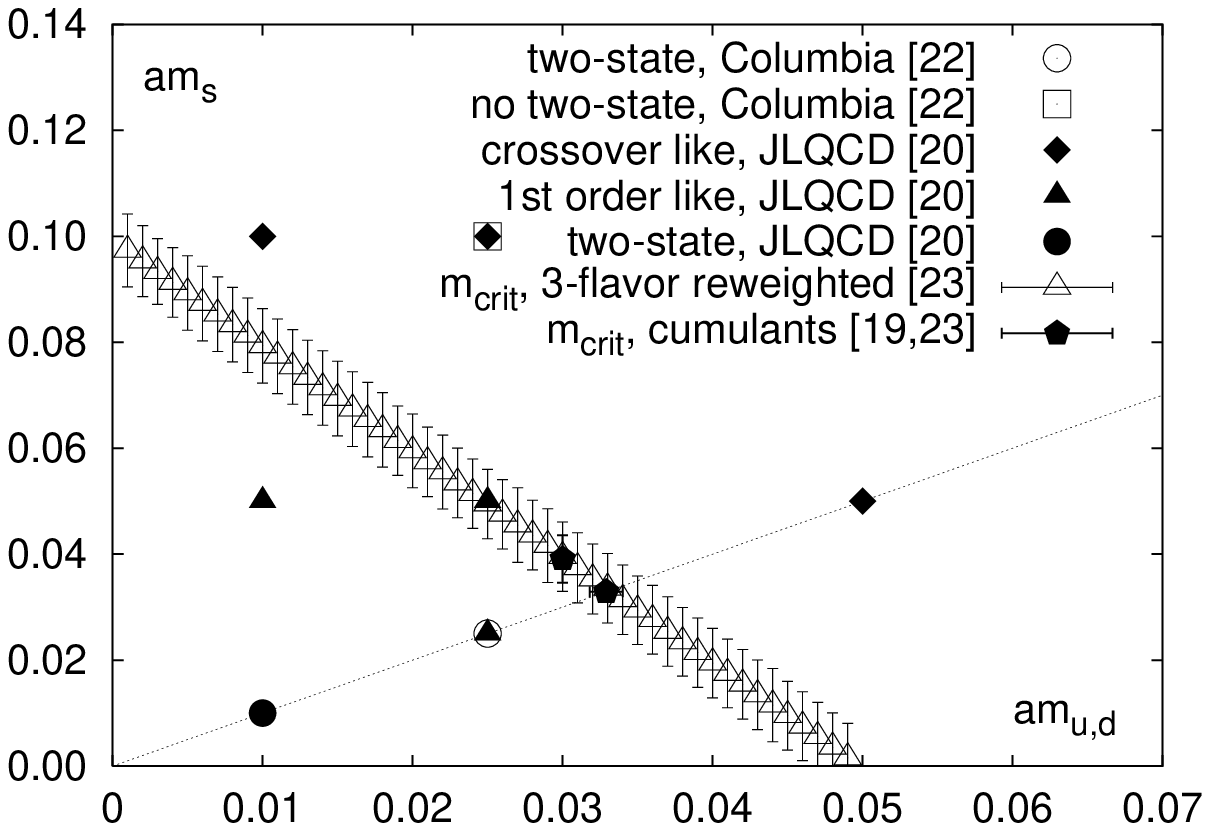,width=72mm}
\end{minipage}
\caption{The phase diagram, expected (left) and lattice
         data (right), in the plane of
         strange and degenerate u,d quark masses.
\label{fig:phase}
         }
\end{figure}

On quite general grounds it is expected that 
at high temperature or baryon density
QCD undergoes
a transition from hadronic matter to the quark-gluon plasma.
However, whether the transition is characterized by a truly
singular behavior of the partition function leading
to a first or second order phase transition, or whether
it merely is a crossover with rapid changes in some observables,
crucially depends on the values of the quark masses.
At a transition 
the details of the microscopic interactions are negligible compared
to the dominant long range correlations, which are determined by
global symmetries. 
Expectations on the nature of the transition
have thus been derived
from studies of simpler systems with the same global
symmetries as QCD \cite{univ}, see Figure \ref{fig:phase}. 

The method to locate a transition usually is to look for
rapid changes of order parameters like the Polykov loop $L$
(in the pure gauge case) or the chiral condensate (in the
chiral limit) and for peaks of their fluctuations, 
i.e.~susceptibilities as e.g.
$\chi_L = V \langle ( L - \langle L \rangle )^2 \rangle$. 
The locations of these changes or peaks define (pseudo-)critical
couplings $\beta_c$, which can be turned into temperatures through
the knowledge of a zero temperature quantity like a hadron
mass $am_H(\beta_c)$ at the same coupling,
$T_c / m_H = 1/(N_t am_H(\beta_c)$.
The nature and critical properties of the transition can be obtained
from scaling dependencies of various quantities on volume or 
external fields like the quark mass.

In the $SU(3)$ pure gauge theory, 
the expected first order transition was verified numerically \cite{su3_crit} 
quite a while ago.
In the chiral limit of two-flavor QCD the relevant global symmetry
group most likely is $SU_R(2) \times SU_L(2) \simeq O(4)$.
In analogy to the $O(4) \; \sigma$ model, one therefore expects \cite{WilPi} 
a second order phase
transition with $O(4)$ critical properties in the chiral limit. 
At non-vanishing quark masses the chiral symmetry is broken explicitely
and a crossover is expected.
All existing lattice calculations 
\cite{fkel,japan_stagg,milc_crit,wilson_crit,engels}
support a continuous transition. However, $O(4)$ values for the
critical exponents could not yet definitively be identified.
Alternatively, the anomalous $U_A(1)$ could become restored
effectively through the disappearance of topologically non-trivial
zero modes of the Dirac matrix at or even below the critical temperature 
rendering the transition first order \cite{WilPi}.
This case would be reflected in a degeneracy \cite{Cohen} 
between the pion and
the scalar isovector meson ($\delta / a_0$), which does not seem to be
realized \cite{Vranas}.

For three degenerate massless quarks, again from symmetry reasons, one expects 
a first order transition. The first order property of the transition 
should extend from the chiral
limit to an end point at some non-zero 
critical quark mass $m^c_q$,
where the transition becomes continuous and is
expected to be in the Ising universality class \cite{Gavin}.
A recent lattice
simulation \cite{ChSchmidt} supports this picture
even with respect to the universality class. 
The location of the end point, however, is
not a universal property and the current estimates
\cite{ChSchmidt,Aoki_3,Liao} vary between 
190 and 260 MeV for the corresponding pseudoscalar mass, depending
on the particular fermion action chosen.

For non-degenerate quarks, i.e. a strange quark heavier than the
two light ones, $m_s \neq m_{u,d}$, the end point becomes an
end line separating the first order from the crossover regime,
see Figure \ref{fig:phase}. In the vicinity of the degenerate three-flavor
end point the slope of the line can be obtained from a Taylor expansion,
$m_s^{c} = m^{c} + 2 (m^{c} - m_{u,d})$. In the currently
accessible quark mass range, this
relation is consistent with all available 
lattice data \cite{ChSchmidt,Aoki_3,columbia,mc},
see Figure \ref{fig:phase}.
If one assumes that such a linear extrapolation holds down to the
physical light quark mass values, the resulting ratio
$m_s^{c} / m_{u,d}^{phys} \simeq 5 - 10$ 
would put
the point of physically realized QCD,
$m_s^{phys} / m_{u,d}^{phys} \simeq 20$ 
(see e.g. \cite{Wittig} for current lattice results),
into the crossover regime of the phase diagram.

The location of the phase transition, i.e.~its  critical temperature,
has been known rather precisely for quite a while in the pure $SU(3)$ gauge
theory \cite{Boyd_eos,Iwa_tc}. 
It has since then been confirmed
by studies which explored the efficiency of various improved
actions \cite{tc_q_us,eos_q_iwa,Necco}, or anisotropic lattices
\cite{tc_q_taro,tc_q_cppacs}, in reducing
finite lattice spacing effects.
The number is most readily given in terms of the dimensionless ratio
to the string tension, $\sqrt{\sigma} \simeq 425 {\rm MeV}$, as
\begin{equation}
T_c / \sqrt{\sigma} = 0.632 \pm 0.002,
\end{equation}
which was obtained as a weighted average over all available lattice data.

\begin{figure}[t]
 \begin{minipage}[t]{.485\linewidth}
\hspace{-5mm} 
 \epsfig{file=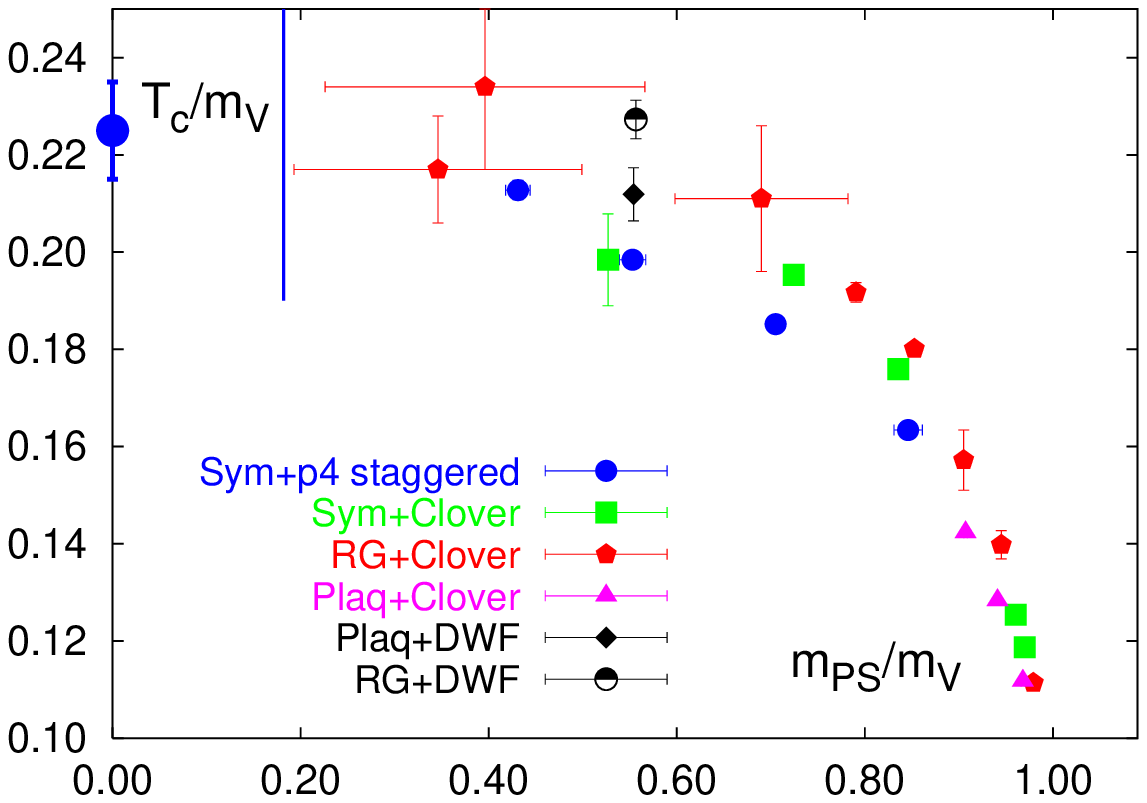,width=64mm}
\end{minipage}
\hfill
 \begin{minipage}[b]{.500\linewidth}
\epsfig{file=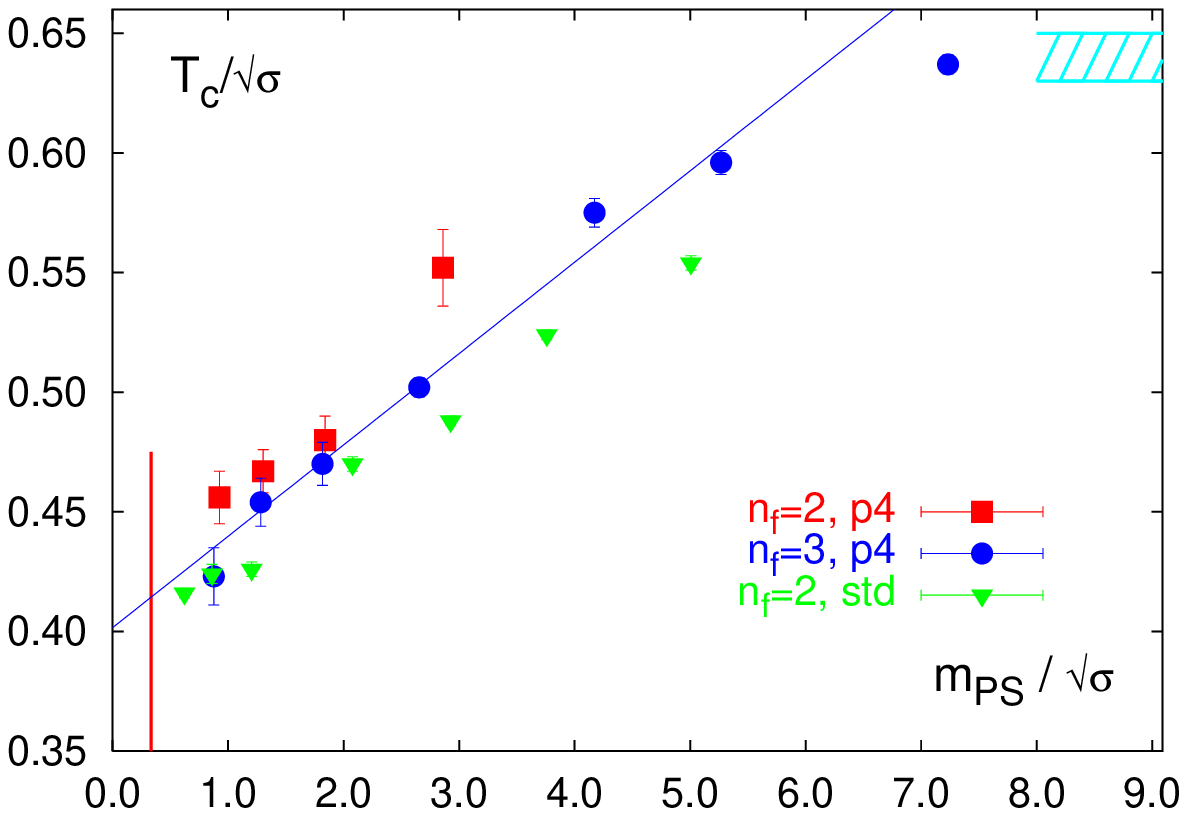,width=64mm}
\end{minipage}
\caption{Left: $T_c$ in units
of the vector meson mass for $N_f=2$, for a variety of improved actions
\protect\cite{Peikert,milc_tc,japan_tc,Urs_tc,dwf_tc}.
Right: $N_f=2,3$ for improved (p4) staggered
quarks, compared to $N_f=2$ standard (std) simulations \protect\cite{Peikert}. 
}
\label{fig:crit_temp}
\end{figure}

For QCD including light dynamical quarks
the current results for the critical temperature are summarized
in Figure \ref{fig:crit_temp}.
The left panel contains 
data \cite{Peikert,milc_tc,japan_tc,Urs_tc,dwf_tc}
for two flavors, all obtained
with improved actions. Discrepancies between Wilson and staggered
quarks seen in earlier computations with standard discretizations
are greatly reduced. $T_c$ is plotted versus the zero-temperature
pseudoscalar mass as a measure of the quark mass, both in units
of the vector meson mass. The strong decrease of the
ratio $T_c/m_V$ with increasing quark mass is due to the rising
vector meson mass which diverges in the infinite quark mass limit,
$m_{PS}/m_V \rightarrow 1$. 
Close to the chiral limit two conflicting
quark mass dependencies should influence the ratio
$T_c/m_V$: the vector mass should increase linearly with the quark mass,
leading to a decrease of the ratio 
with $\sqrt{m_{PS}}$.
This is counteracted by the quark mass dependence of $T_c$ itself,
rising proportional to
$m_q^{1/\beta \delta}$, which amounts to $\sim m_{PS}^{1.1}$ for 
$O(4)$ values for the critical exponents. 
Thus, at small enough
quark and pseudoscalar mass, respectively, a linear decrease towards the chiral
limit is expected to win. This effect is not yet seen.
A scale for $T_c$ much less affected by the quark mass 
is the string tension
which is used on the right panel of Figure \ref{fig:crit_temp}.
Indeed, this figure shows that the critical temperature is decreasing
with decreasing quark mass. The decrease appears to be compatible
with a linear dependence on the pseudoscalar mass as expected for
$N_f=2$. The linear dependence, however, also seems to hold for
three degenerate quarks.
Note also, that $T_c$ 
is reduced half way down from the quenched value, shown as the band
to the very right of the figure,
already at a pseudoscalar mass 
as large as 1500 MeV.
It is thus not the chiral dynamics which controls the critical 
temperature, and one may speculate that a resonance gas picture
is more appropriate to describe the thermodynamics close to $T_c$.
The overall quark mass dependence for both two and three flavors
is rather weak.
The difference between $N_f=2,3$ is small and amounts to about
20 MeV, independent of the quark mass.
Extrapolating the results to the chiral limit, 
one obtains
\begin{eqnarray}
\underline{\rm 2-flavor~ QCD:} &&
T_c  = \cases{
(171\pm 4)\; {\rm MeV}, & clover-improved Wilson \nonumber \cr
   ~& fermions\cite{japan_tc}     \nonumber \cr
(173\pm 8)\; {\rm MeV}, & improved staggered     \nonumber \cr
   ~& fermions\cite{Peikert}     \nonumber \cr
             }  \nonumber \\
\underline{\rm 3-flavor~ QCD:} &&
T_c  = \; \; \;\; (154\pm 8)\; {\rm MeV}, \; \; \hspace*{0.1cm}
     \mbox{improved  staggered} \nonumber \cr
& & \hspace*{4.6cm} \mbox{fermions\cite{Peikert}}
\nonumber
\end{eqnarray}
where $m_\rho$ has been used to set the scale.
Although the agreement between staggered and Wilson quarks
is striking, one should keep in mind that
the errors are statistical only and do not account for the 
as yet unknown systematic errors originating from 
a non-vanishing lattice spacing.
One might hope, though, that these are small
since improved actions have been used in
both simulations.

\section{Equation of State}

Energy density $\epsilon (T)$ and pressure $p(T)$ as a function
of temperature are certainly among the most
fundamental thermodynamic quantities of the quark gluon plasma
governing, e.g., the time evolution
of the plasma once being created in heavy ion collisions.
At high temperature the relevant partonic degrees of freedom
have momenta of order $\pi T$ which are
strongly affected by the
UV cut-off introduced through the finite lattice spacing.
It is thus important to gain control over the discretization effects
and carry out the continuum limit $a \rightarrow 0$.
Again, this was first accomplished in the pure gluon theory
\cite{Boyd_eos}, later confirmed in studies utilizing improved
actions and/or anisotropic lattices \cite{tc_q_us,eos_q_iwa,tc_q_cppacs}.
As calculations in the ideal gas limit of lattice perturbation theory reveal,
the discretization effect is even more
pronounced in the presence of dynamical quarks. 
Since on the other hand the signal for these
quantities vanishes proportional to $a^4$
the use of improved actions,
which are designed to reduce the UV cut-off effects, seems mandatory.
Moreover, the
simulations in the quenched approximation have shown that this
improvement helps to extract the continuum limit also in the
intermediate $T$ range between 2 and 4 $T_c$.

\begin{figure}[t]
 \begin{minipage}[t]{.485\linewidth}
\epsfig{file=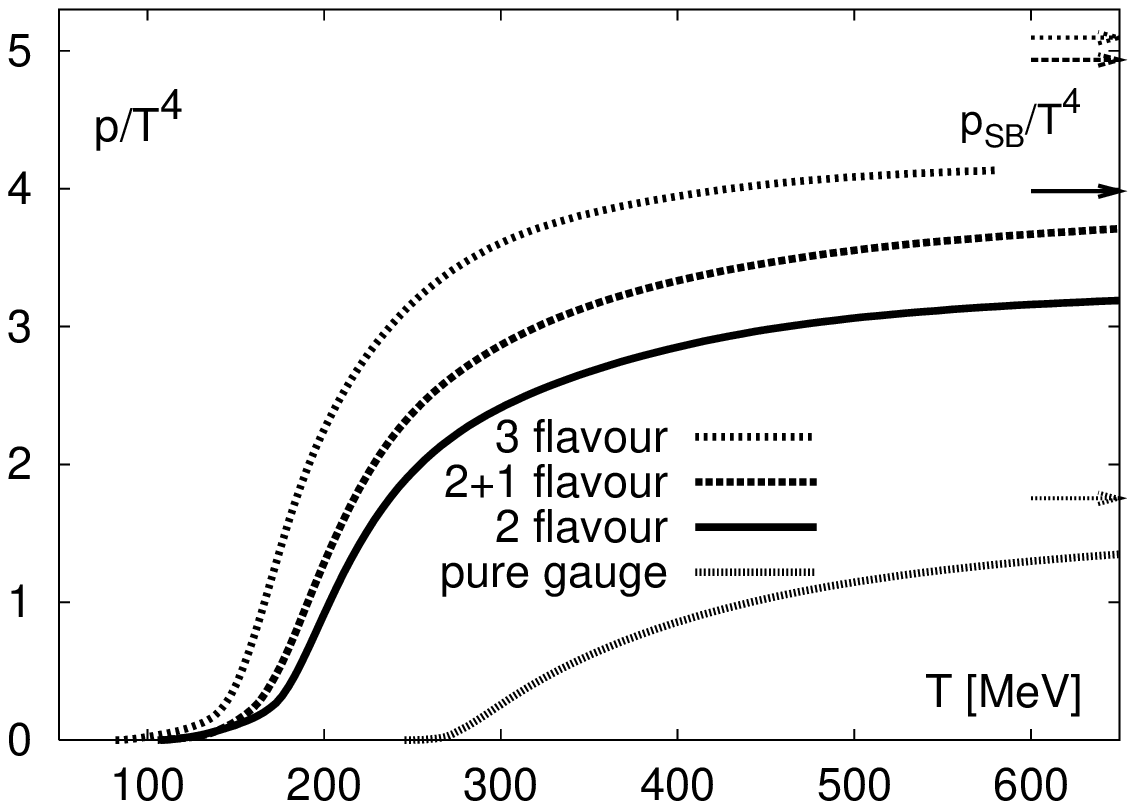,width=62mm}
\end{minipage}
\hfill
 \begin{minipage}[b]{.485\linewidth}
\epsfig{bbllx=25,bblly=37,bburx=533,bbury=430,
       file=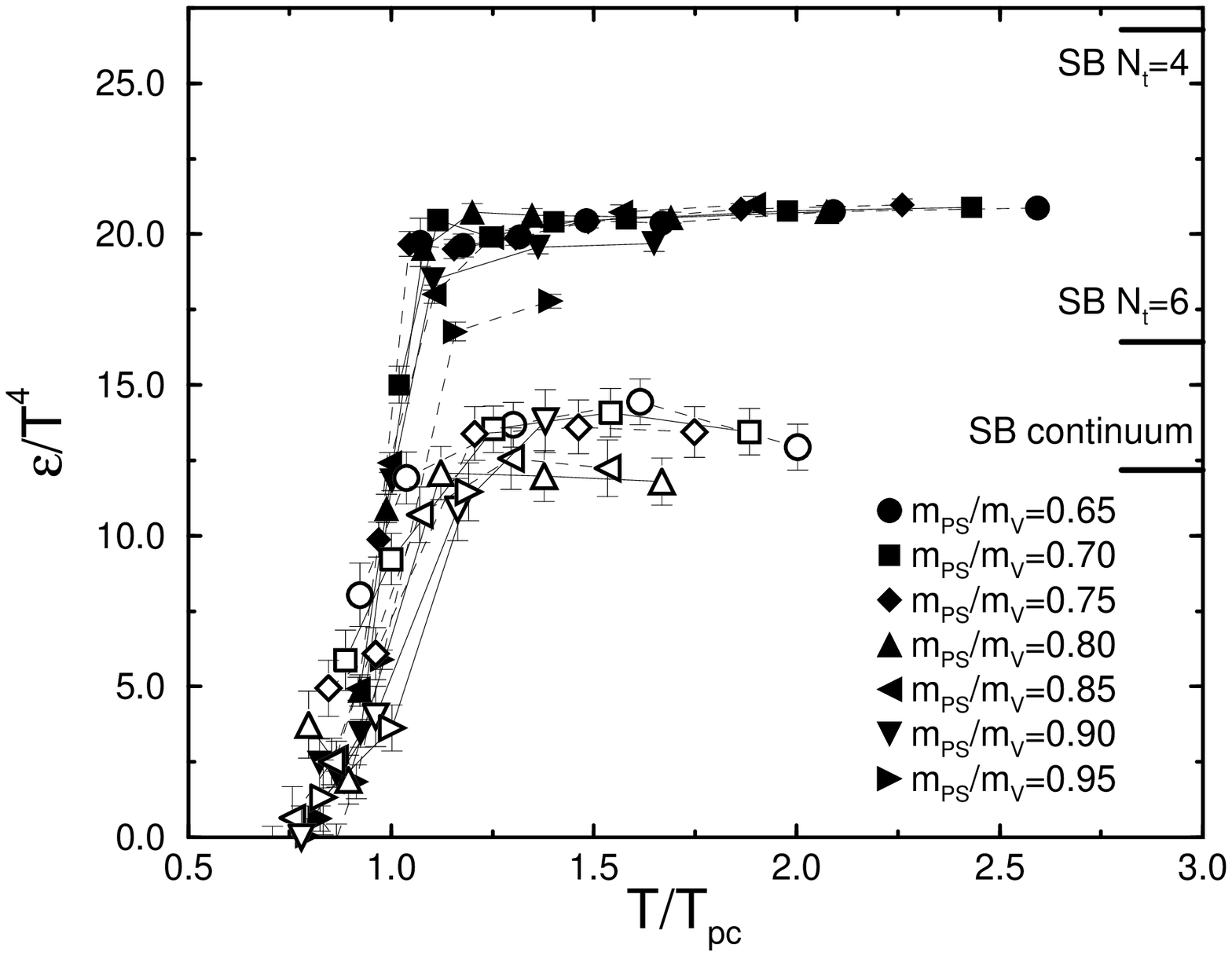,height=4.50cm,width=6.4cm}
\end{minipage}
\caption{
Left: flavor dependence of the pressure 
for $N_t=4$ lattices compared to a continuum extrapolated
pure gauge result \protect\cite{Peikert_eos}. 
Right: energy density for $N_f=2$ improved Wilson quarks on $N_t=4$
(filled symbols)  and $N_t = 6$
(open symbols) \protect\cite{Wilson_eos}. Marks on the right side
denote Stefan-Boltzmann limits.
\label{fig:pressure}}
\end{figure}

The results of a computation of the pressure with an improved action
in the gauge as well as in the fermion 
sector \cite{Peikert_eos}
are shown in 
Figure \ref{fig:pressure}. The data have been obtained for 
$N_f=2,3$ with (bare) mass $m_q/T = 0.4$ as well as
for $N_f=2+1$ with a heavier mass $m^s_q/T = 1$.
For comparison the figure includes the
continuum extrapolated quenched result. 
The figure shows a rapid rise of the pressure in a transition region.
The critical temperature as well as the
magnitude of $p/T^4$ reflect the number of degrees of
freedom liberated at the transition.
In fact, when the pressure is normalized to the appropriate 
Stefan-Boltzmann limits shown as arrows in the figure, the function
$p(T/T_c)/p_{SB}$ turns out to be almost flavor independent.
It is thus well described by
\be
\frac{p}{T^4} = \left( 16 + \frac{21}{2} N_f \right) \frac{\pi^2}{90} \times f(T/T_c),
\ee
where 
the ideal gas limit only is modified by an
apparently flavor independent function $f(T/T_c)$.
At this stage it should be pointed out, though, that the full QCD results 
have not yet been extrapolated to the
continuum limit. However, the experience gathered perturbatively
and in the quenched approximation leads one to expect that 
the finite $a$ effects will distort these findings by not more 
than 10\%.

The most recent results on the energy density are shown in
Figure \ref{fig:pressure} to the right.
Note that the dependence on the quark mass appears to be rather
weak over a wide range of values.
The data has been obtained for $N_f=2$ improved Wilson quarks
\cite{Wilson_eos} on $N_t =4,6$ lattices.
However, in the particular discretization chosen
the improvement does not reduce the infinite temperature
finite lattice spacing effects.
This is visible in the figure by the big difference
in the Stefan-Boltzmann limits computed on the two lattices with
different $N_t$ values.
On the other hand, the cut-off effects are not important close to $T_c$
as a comparison between $N_t = 4$ and 6 results reveals.
This is expected since in this temperature regime
the correlation length is large and infrared modes
dominate. Thus, a comparison of
the different fermion discretizations is meaningful.
In fact, improved staggered data are
available \cite{Peikert_diss} and lead
to a consistent estimate for the critical energy density
of $\epsilon (T_c) \simeq (6 \pm 2) T_c^4$,
which also agrees with the value originating from an earlier simulation
using the standard staggered action \cite{MILC_eos}.

\begin{figure}[t]
\centerline{\epsfxsize=6cm
\epsfbox{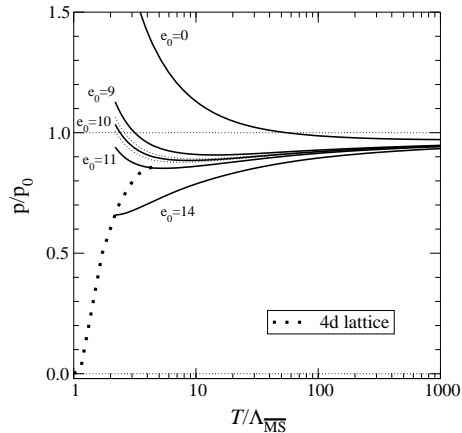}}
\caption[a]{The pressure of pure gauge theory from dimensional
reduction, with an as yet undetermined constant $e_0$. From \cite{3pres}.}
\label{fig:3dp}
\end{figure}

In all cases,
up to the highest temperature investigated,
pressure and energy deviate 
substantially from the ideal gas behavior.
The deviation
is too big to be reproduced in ordinary high temperature
perturbation theory which converges badly in this temperature range
\cite{Zhai}.
A more refined technique as hard thermal
loop resummation \cite{Blaizot} gets closer to the lattice results. 
The question of the approach to the ideal gas limit at higher 
temperatures can 
also 
be addressed by dimensional reduction,
which splits the thermodynamic functions into perturbatively calculable
parts and a remainder to be simulated in 
the effective theory \eq (\ref{actc}) \cite{braaten}.
Unfortunately, this requires determination of 
an integration constant $e_0$ by a four-loop calculation, which has not
been completed yet. However, results \cite{3pres} treating $e_0$ like a free parameter
and matching it to the 4d simulations are shown in Figure \ref{fig:3dp}.
If the eventual value for $e_0$ lies in the right range, the whole 
picture makes sense
and explains how ideal gas values are only attained 
at asymptotic temperatures.

\section{Screening Masses}
\label{sec:screen}

Essentially all static equilibrium properties of a thermal 
quantum field theory are encoded in its equal time correlation functions.
Even though these quantities may not be 
directly accessible in heavy ion collision experiments, their theoretical knowledge
provides us with the relevant dynamical length scales in the plasma, from which
conclusions about the acting degrees of freedom and their physical effects may be drawn.
The connected spatial correlation functions of gauge-invariant,
local operators $A(x)$,
\be
C(|{\bf x}|)=\langle A({\bf x})A(0)\rangle_c\sim \ex^{-M|\bfx|},
\label{eq:corrfct}  
\ee
fall off exponentially with distance. The
``screening masses'' $M$ 
are the eigenvalues
of the spacewise transfer matrix of the corresponding
lattice field theory, and classified by its symmetries.
Because of the shortening of the euclidean time direction at $T>0$, the continuum 
rotation symmetry of the hypertorus orthogonal to the correlation direction is broken down 
from $O(3)$ to $O(2)\times Z(2)$, and its appropriate subgroup for the
lattice theory is $D_h^4$. The irreducible representations and the classification
of operators have been worked out for pure gauge theory \cite{gros, dg} as well as for
staggered quarks \cite{symg}. 
Physically, the screening masses correspond to the inverse length scale over 
which the equilibrated medium is sensitive
to the insertion of a static source carrying the quantum numbers of $A$. Beyond
$1/M$, the source is screened and the plasma appears undisturbed.
Technically, the computation of these quantities is equivalent to 
spectrum calculations at zero temperature.

\subsection{Hadronic Screening Masses}

\begin{figure}
\leavevmode
\hspace*{-0.2cm}
\epsfysize=5cm
\epsfbox{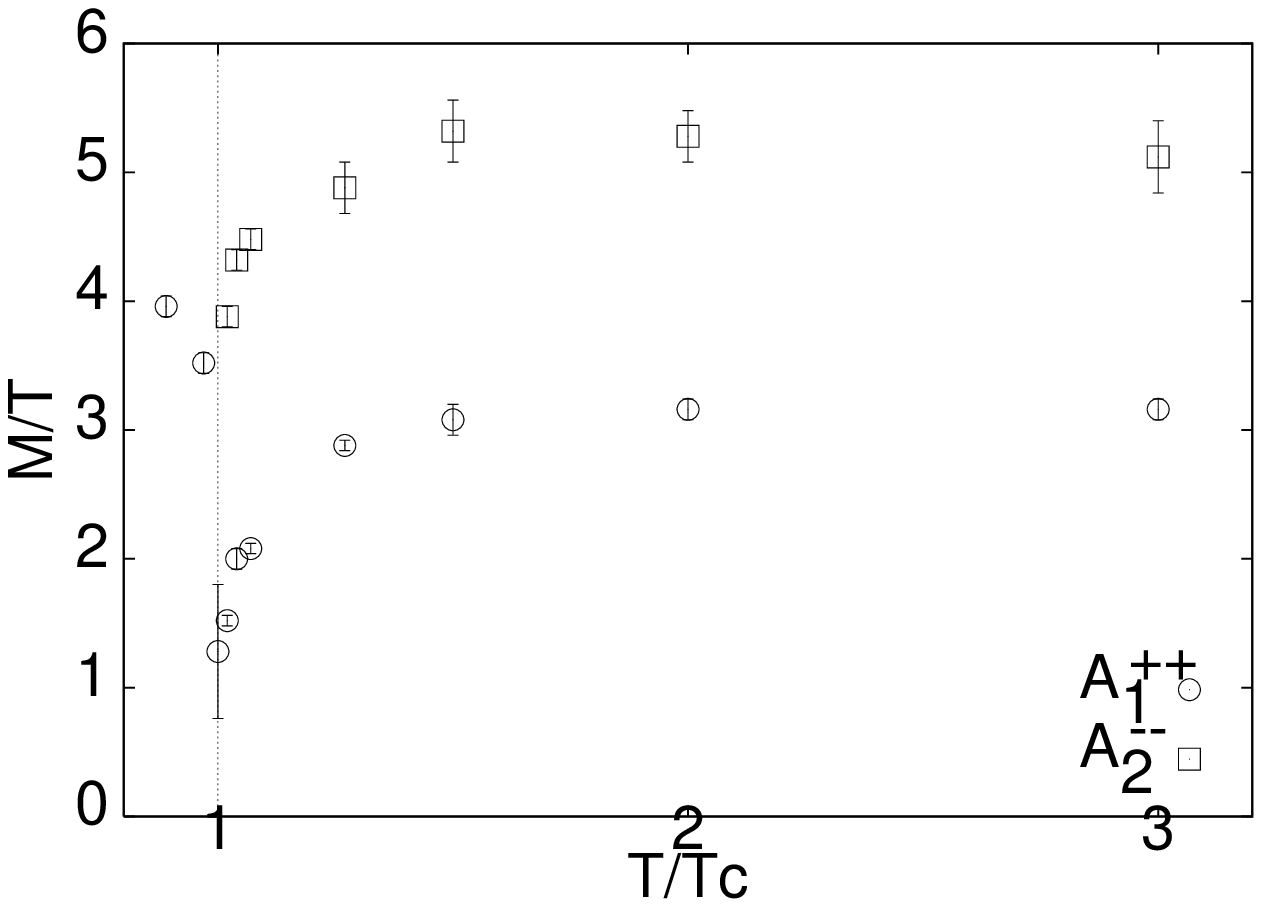}
\epsfysize=5.1cm
\epsfbox{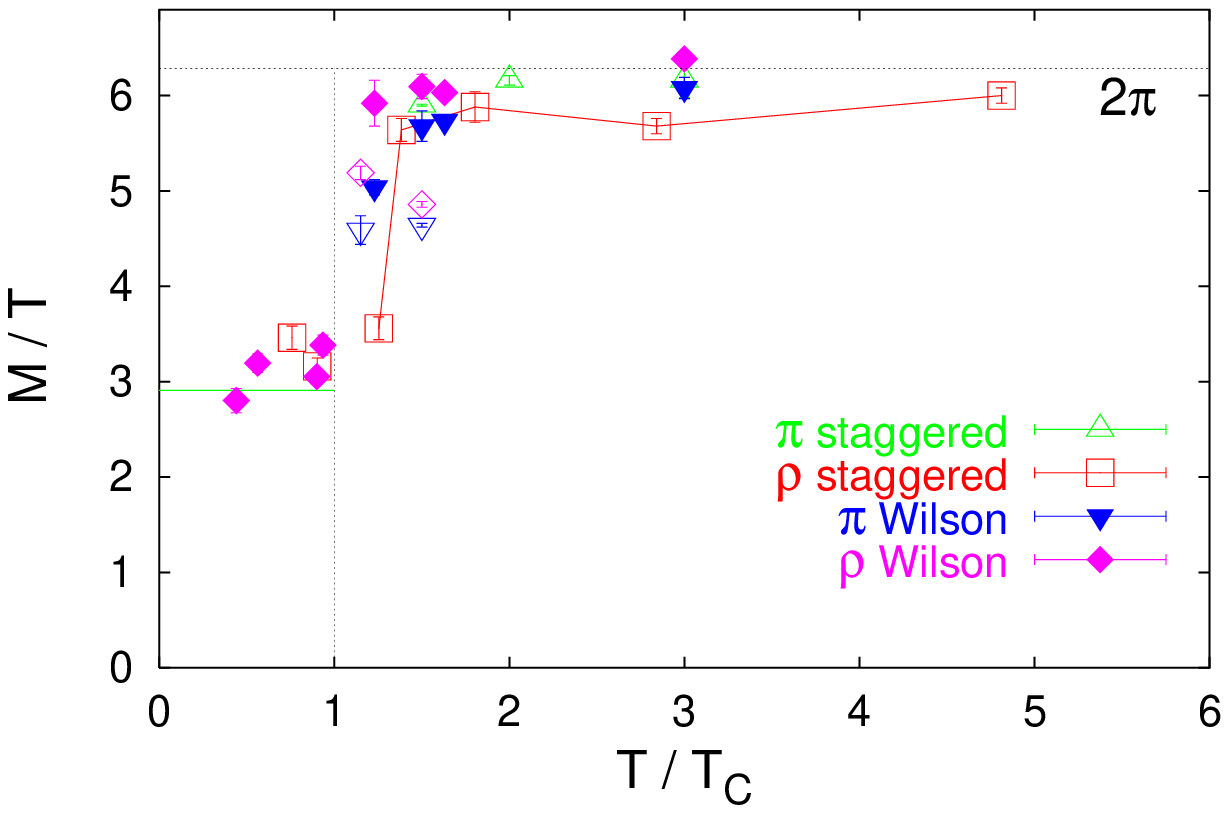}
\caption[]{Left: Screening masses for the pure gauge theory,
corresponding to the continuum $0^{++}_+$ (circles) and 
$0^{+-}_-$ (squares) channels. From \cite{dg2}.
Right: Mesonic screening masses in the quenched chiral limit.
Below $T_c$, $M_{\rho}/T_c$ \protect\cite{PSchmidt,Binew} is plotted,
the line denotes the $T=0$ value. Open Wilson symbols denote anisotropic
lattices \protect\cite{Nucu_1}. Staggered pion data are extrapolated
to $a=0$ \protect\cite{Sourendu2}, Wilson and staggered rho are from
$N_t=8$ \protect\cite{PSchmidt} and 16 \protect\cite{Binew} lattices.
The free quark limit has not been corrected for finite
         volume effects.
}
\label{screen}
\end{figure}

\begin{figure}[t]
\vspace*{-1.75cm}
\epsfysize=5cm
\leavevmode
\epsfbox{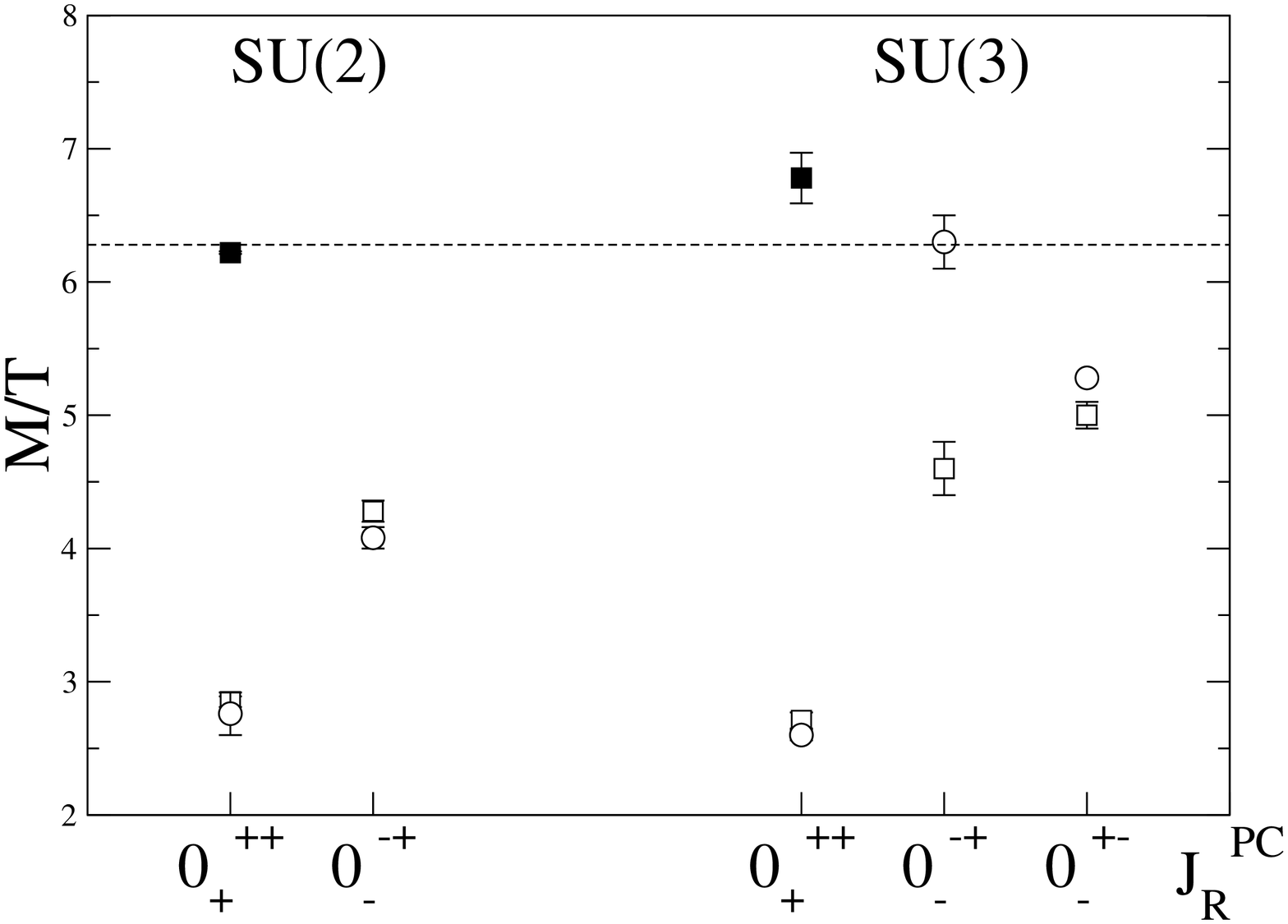}
\epsfysize=7cm
\epsfbox{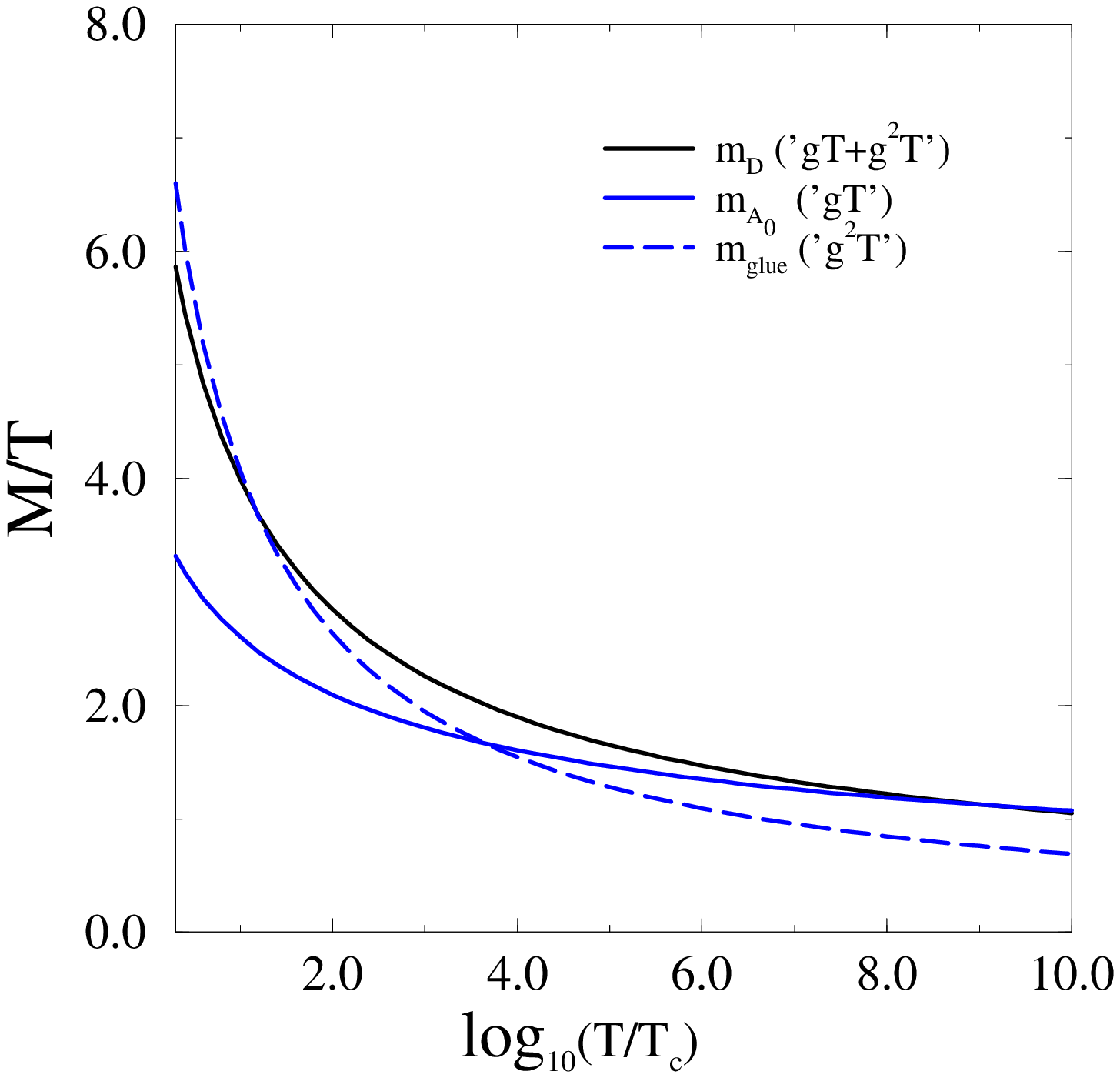}
\caption[]{Left: Pure gauge screening masses
at $2 T_c$ from simulations of the 
4d (circles) \cite{dg} and the
dimensionally reduced \cite{hp,hlp1} theories, where
filled squares correspond to $0_+^{++}:\tr (F_{12}^2):m_{glue}$,
open squares represent 
$0^{++}_+:\tr (A_0^2):m_{A_0}, 0^{-+}_-:\tr (A_0F_{12}):m_D$ 
and $0^{+-}_-:\tr (A_0^3)$.
Right: Temperature dependence of the two lowest $0^{++}_+$
and the lowest $0^{-+}_-$ states \cite{hlp1,mik}.
}
\label{drcomp}
\end{figure}

Figure \ref{screen} (left) shows results for the lowest lying screening masses 
corresponding to glueball operators around $T_c$.
Comparison with dimensionally reduced results 
(cf.~Figure \ref{drcomp}) shows 
that the pure gauge results are close to continuum physics. 
In the range $0.8 T_c<T<T_c$, the lowest scalar screening mass is observed to be roughly 20\%
lower than the lightest scalar glueball at zero temperature, $M(T)/M_G(T=0)\sim 0.8$.
At $T_c$ a sharp dip is observed, after which the screening masses appear 
to be proportional to $T$.
Screening states with larger masses show the same qualitative behavior above $T_c$.

Apart from operators $A$ made from pure glue in
Eq.~(\ref{eq:corrfct}),
also spatial correlations of meson operators 
have been investigated, both in the quenched approximation
as with various numbers of dynamical fermions
\cite{DeTarKogut,Nucu_400,PSchmidt,Binew,
HT_ua1,Nucu_1,gupta_lacaze,Sourendu2,Lagae}.
The picture which has emerged so far
is illustrated by some of the available data
shown in Figure \ref{screen} (right).
Below $T_c$, 
the screening masses have shown neither a marked
temperature dependence nor, correspondingly,
a drastic difference to zero temperature masses.
At temperatures above $T_c$ 
spatial (as well as temporal) correlation functions 
reflect the
restoration of the chiral $SU_L(N_f) \times SU_R(N_f)$ symmetry.
In particular, the vector and axial vector channel become 
degenerate
independent of the discretization and of the number of dynamical
flavors being simulated.
Moreover, the pion ceases to be a Goldstone boson
and acquires a screening mass.
A degeneracy is further observed within errors in the pseudoscalar
and isoscalar scalar ($\sigma / f_0$) channel although the latter, 
for technical reasons, 
is difficult to access on the lattice. Nevertheless, 
screening masses obtained from fits to correlation functions 
\cite{Lagae} and  
susceptibilities (at finite lattice spacing) \cite{MILC_ua1}
lead to a consistent picture.
At $T_c$ the anomalous $U_A(1)$ most likely is not yet
restored effectively \cite{Vranas}, 
see also \cite{HT_ua1,Lagae,MILC_ua1,Christ},
as has been mentioned already in Section~\ref{sec:phase}.

At high temperature, the screening masses are expected to approach 
the free quark propagation limit, $M \rightarrow 2 \pi T$.
In fact, already at temperatures of about 1.5 $T_c$,
the results for the vector channel are not far from this value.
If the finite volume effects for free quark propagation are
taken into account, the
deviations amount to about 15 \% and decrease only slowly
with temperature.
Moreover, at temperatures $\gsim 1.5 T_c$, 
in the Wilson discretization
pion and rho become nearly degenerate
already at finite lattice spacing.
Earlier staggered simulations had found discrepancies
here.
However, 
a recent paper \cite{Sourendu2} reports that also
in the staggered discretization a degeneracy 
of pseudo-scalar and vector is reached in the
continuum limit. 
Quenching effects are found to be below 5\% for $T>T_c$ \cite{ggm}.

In the plasma phase, screening masses can also be investigated within the dimensionally
reduced theory, cf.~Section \ref{sec:dr}.
In this framework they correspond to the spectrum of the transfer matrix for the 3d theory.
The associated Hamiltonian respects
SO(2) planar rotations, two-dimensional parity $P$,
$A_0$-reflections $R$, and the symmetry is again
$SO(2)\times Z(2)\times Z(2)=O(2)\times Z(2)$. Remember however that in this
setup one is interested in soft modes, while the Matsubara frequency
$\sim 2\pi T$ represents the UV cut-off for the effective theory.

Figure \ref{drcomp} displays numerical results for the lowest
screening masses for the SU(2) and SU(3) pure gauge
theories, obtained through simulations of both the 3d effective theories 
and the full 4d theories, in various quantum number channels.
Even at a temperature as low as $T\sim 2T_c$, good agreement 
is observed except in one case. 
Note that the error bars are only statistical,
while both calculations have systematic errors: 
the effective theory has truncation errors, while
the 4d simulations are not infinite volume and 
continuum extrapolated. Estimating such effects at 10-15\%, 
we conclude that within current accuracy dimensional reduction gives a quantitative
description of screening lengths for temperatures $T\gsim 2T_c$. The same conclusion is reached
with other observables like the spatial string tension \cite{bali}, static 
potentials \cite{rold} or gauge fixed propagators (cf.~Section \ref{sec:gluon}).
For larger temperatures the $M/T$ get lower and the spectrum denser, as shown in
Figure \ref{drcomp} (right) for a few states. A large number of such screening masses
in many quantum number channels for various numbers of flavors has been computed in 
\cite{hlp1}.

\subsection{\label{hpt}How Perturbative is the Plasma?}

The approach of dimensional reduction is particularly valuable in disentangling
contributions from different degrees of freedom, thanks to accurate mixing analyses,
as well as for treating larger temperatures $T\gg T_c$ which cannot be reached by 
4d lattices. This can be used to inspect to what extent the plasma behaves perturbatively.
Figure \ref{drcomp} then teaches us that for any reasonable temperature
the largest correlation length of gauge-invariant operators belongs to
the $A_0$ degrees of freedom and not to the $A_i$, in contrast to the naive parametric
ordering, \eq (\ref{eq:scale}). Only asymptotically is the perturbative ordering restored.

Of particular phenomenological relevance for the QCD plasma is the Debye
mass, whose inverse gives the length scale over which colour-electric
flux is screened. Its expansion in powers
of coupling constants reads
\be \label{debye}
m_D = m_D^{\rm LO}+{Ng^2T\over4\pi}\ln{m_D^{\rm LO}\over
g^2T} +
c_N g^2T + {\cal O}(g^3T),
\ee
with the leading order perturbative result $m_D^{\rm LO}=(N/3+N_f/6)^{1/2}gT$.
At next-to-leading order $\sim g^2T$, only a logarithm can
be extracted~\cite{rebhan}, whereas the coefficient $c_N$ is entirely
non-perturbative.

A gauge invariant definition of the Debye mass
has been suggested in \cite{ay},
according to which it corresponds to the mass
of the lightest gauge invariant screening state odd under euclidean time reflections.
In the 3d effective theory the latter is replaced by the scalar reflection
$R$, and the Debye mass thus corresponds to the
$0^{-+}_-$ ground state $\sim \tr A_0 F_{12}$.

The coefficient $c_N$ can
be measured separately from the exponential
decay of a Wilson line in a 3d pure gauge theory~\cite{ay},
which has been performed in \cite{lp}.
This allows to disentangle the contributions from different scales,
as shown in Table~\ref{deb}. The
${\cal O}(g^3T)$ corrections
are less than 30\% even at temperatures as low as $T=2T_c$,
and they disappear entirely for asymptotically large temperatures.
On the other hand, the dominant scale
is again $\sim g^2 T$ for all temperatures of interest,
in contrast to the naive expectation $\sim gT$. Only at asymptotically
high temperatures is the perturbative picture restored.
These findings are compatible with and explain the behavior of
the pressure density, which attains its ideal gas limit only at asymptotic temperatures.
There the effect is less pronounced, since the pressure is dominated by hard modes and 
soft modes provide corrections, whereas 
the screening masses discussed here couple exclusively to soft modes.

\begin{table}
\caption[]{Contributions of the first two, the third and higher order terms in \eq (\ref{debye})
to the total $m_D$ \cite{hlp1}.}
\begin{center}
\begin{tabular}{@{}ccccc@{}}
\hline\hline
SU(3), {\small $N_f=0$} &$m_D/g^2T$ &
$(1+2)$ &
$c_3$ &
${\cal O}(g^3T)/(g^2T)$ \\
\hline
$T=2T_c$  & 1.70 (5) & 0.514 & 1.65(6) & -0.46(6) \\
$T=10^{11}T_c$  & 3.82 (12) & 2.165 & 1.65(6) & 0.00(12) \\
\hline
\end{tabular}
\end{center}
\label{deb}
\end{table}

\subsection{\label{sec:gluon}Gluon Screening Masses}

With energy and pressure densities as well
as the quark number susceptibilities rising rapidly
across the transition, it seems clear that some light constituents 
are liberated in the deconfined phase. 
A longstanding question concerns the nature of these
degrees of freedom and their properties. 
Moreover, 
spatial
correlators of singlet operators
as discussed in the previous subsections, 
and in particular a Debye mass defined
by one of them, are not directly related to phenomena
like $J/\psi$-suppression. These are caused by color charged intermediate states, i.e.~some
kind of constituent.
In perturbation theory one therefore defines the Debye mass in analogy to QED as the 
pole in the $A_0$-propagator \cite{rebhan}, with a perturbative expansion as in 
\eq (\ref{debye}), 
which can be proved to be gauge invariant order by order \cite{kkr}.
On the other hand, at moderate temperatures $T>T_c$
interactions are still strong, and a purely perturbative parton picture
is surely not appropriate. 
Similarly, the propagator of magnetic fields $A_i$ is expected to develop a magnetic mass
on the scale $\sim g^2 T$, which however even at asymptotic temperatures is entirely
non-perturbative \cite{linde}.

Unfortunately, lattice studies of gluon propagators \cite{og} are hampered by several problems.
It is difficult to fix a gauge uniquely and avoid the problem of Gribov copies \cite{grib}.
Moreover, most complete gauge fixings, like Landau gauge, violate the positivity
of the transfer matrix, thus obstructing a quantum mechanical interpretation of the results.
However, recently it was shown analytically \cite{op1}
that gluon correlators can be made gauge invariant
by dressing them with appropriate (gauge fixing) functionals. These correlators
fall off exponentially with eigenvalues of the Kogut-Susskind 
Hamiltonian in the presence of sources \cite{op1}, thus probing the energy of a gluon 
coupled to sources. The spectrum is gauge invariant, while
matrix elements depend on the particular functionals chosen.
This is true for all gauge fixings which are local
in time (to preserve the transfer matrix), such as the Coulomb gauge.

\begin{figure}
\begin{center}
\epsfysize=5cm
\epsfbox{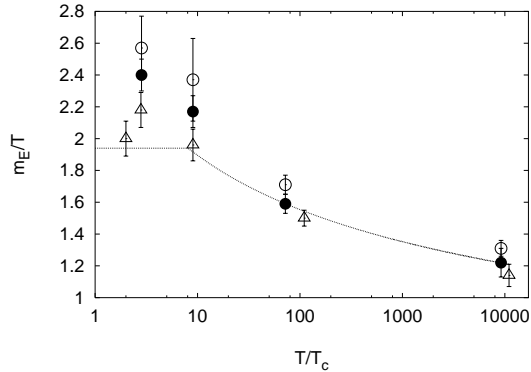}
\end{center} 
\vspace*{-0.5cm}
\caption[]{The electric screening (Debye) mass
in $SU(2)$ gauge theory. 
The line represents a fit to
4d simulations \cite{hkr}, the symbols show simulations of the dim.~red.~theory with different
matching procedures. Triangles correspond to two-loop perturbative matching. From \cite{ckp}.
}
\label{el}
\end{figure}

Numerically, in detail
only SU(2) pure gauge theory has been investigated so far.
Figure \ref{el} shows results obtained for the electric screening mass obtained both in 
4d simulations as well as in simulations of the dimensionally reduced theory, where a variety
of gauges was found to agree with the Coulomb gauge results.
Like for the hadronic screening masses,
with two-loop matching
good agreement between full and effective theory 
for temperatures down to $T\approx 2T_c$ is observed.
Equally consistent is the finding that the leading $\sim gT$ contribution to the
electric mass is subdominant for
temperatures up to ${\cal O}(10^4T_c)$ \cite{hkr}.

Extracting the magnetic mass appears to be more difficult technically. In Landau type gauges,
where positivity is violated, severe finite volume effects are observed \cite{hkr,ckp}. In 
Coulomb type gauges the spatial non-locality of the fixing functionals leads to an exclusive
projection on torelonic states \cite{op1}. However, by introducing explicit local fields to
mimic static sources and extrapolating to their infinite mass limit, the magnetic mass has been
calculated in the high temperature limit, where the dimensionally reduced model is just
the 3d SU(2) pure gauge theory, and the result is \cite{op2}
\be
m_M= 0.36(2) g^2 T.
\ee
It is interesting to note that this compares well with various
resummation methods leading to self-consistent gap equations,
which at one loop yield pole masses in the range of $m_M\sim 0.25-0.38 g^2T$ \cite{mm},
while a two-loop calculation gives $m_M=0.34 g^2T$ \cite{eb}.
Note that these partonic screening masses are much lower than those of the singlet operators,
and hence correspond to larger correlation lengths.
This should play an important role for fragmentation phenomena in the plasma.

For a constituent picture of the plasma to be appropriate, it should be possible
to interpret the singlet operators as multigluon states \cite{gros}. 
For high $T$ within the 3d theory, this means \cite{bp2} that the masses of the
lowest dimensional operators should be approximated by appropriate multiples of $m_E,m_M$. 
This appears to work up to about 10\% for temperatures $T\gsim 100T_c$ \cite{bp2,petr}, 
but accuracy decreases with $T$. E.g.~the two masses shown in 
Figure~\ref{screen} (left) seem to slowly approach their multigluon ratio 3/2 
from above~\cite{dg2}.

\section{Free Energy of Static Quarks}

A different source of information on gluonic excitations
is provided by heavy quark free energies, in particular
of a quark-antiquark pair. 
Such systems are also of interest for the physics of
heavy quarkonia in the medium,
in particular for the question 
of $J/\psi$-suppression \cite{Matsui}.
The $Q\bar{Q}$ free energy is defined \cite{McLerran}
by the partition function
of the thermal heat bath containing a static quark and
antiquark source at separation $\vec R$,
\begin{equation}
\langle \; {\rm tr} L(\vec R) \; {\rm tr} L^\dagger(0) \; \rangle
= \exp\{- (F_{Q \bar Q} \,(|\vec R |,T) - F_0(T)\;)/T\},
\label{eq:poly_corrfct}
\end{equation}
where ${\rm tr} L(\vec x) = (1/N_c) {\rm tr}
\prod_\tau U_\tau(\vec x,\tau)$ is the Polyakov loop
and $F_0$ denotes the free energy of the heat bath.
The gauge invariant Polyakov loop correlation 
averages over color singlet and octet contributions, and the free energy
is a superposition \cite{McLerran, Nadkarni,Nadkarni1},
\begin{equation}
e^{-F_{Q \bar Q}(R,T)/T } =
\frac{1}{9} \;  e^{- F_{1}(R,T)/T } +
\frac{8}{9}   \; e^{- F_{8}(R,T)/T }.
\label{eq:average}
\end{equation}
At zero temperature the free energies reduce to the
heavy quark potentials, 
and at non-zero temperature
they 
also 
exhibit modifications of the potentials by Boltzmann
weighted thermal excitations.

\begin{figure}[t]
\leavevmode
\epsfig{file=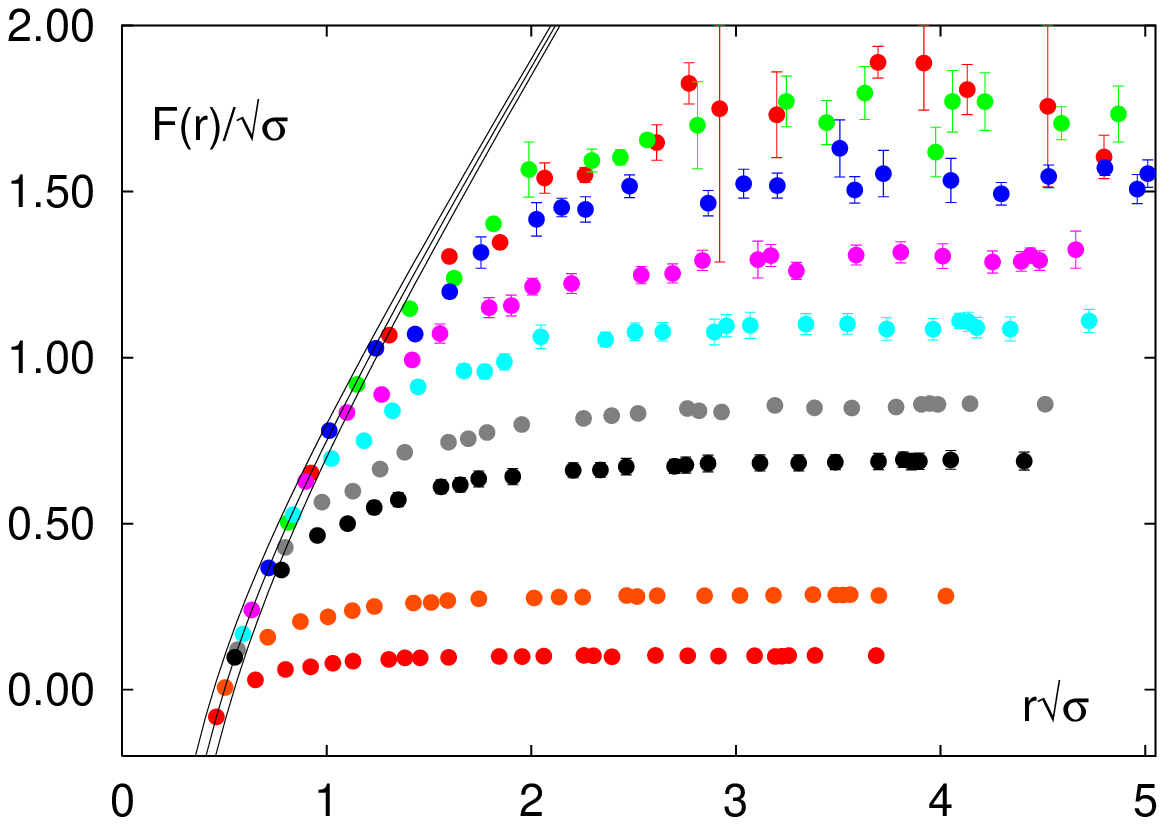,width=70mm}
\epsfig{file=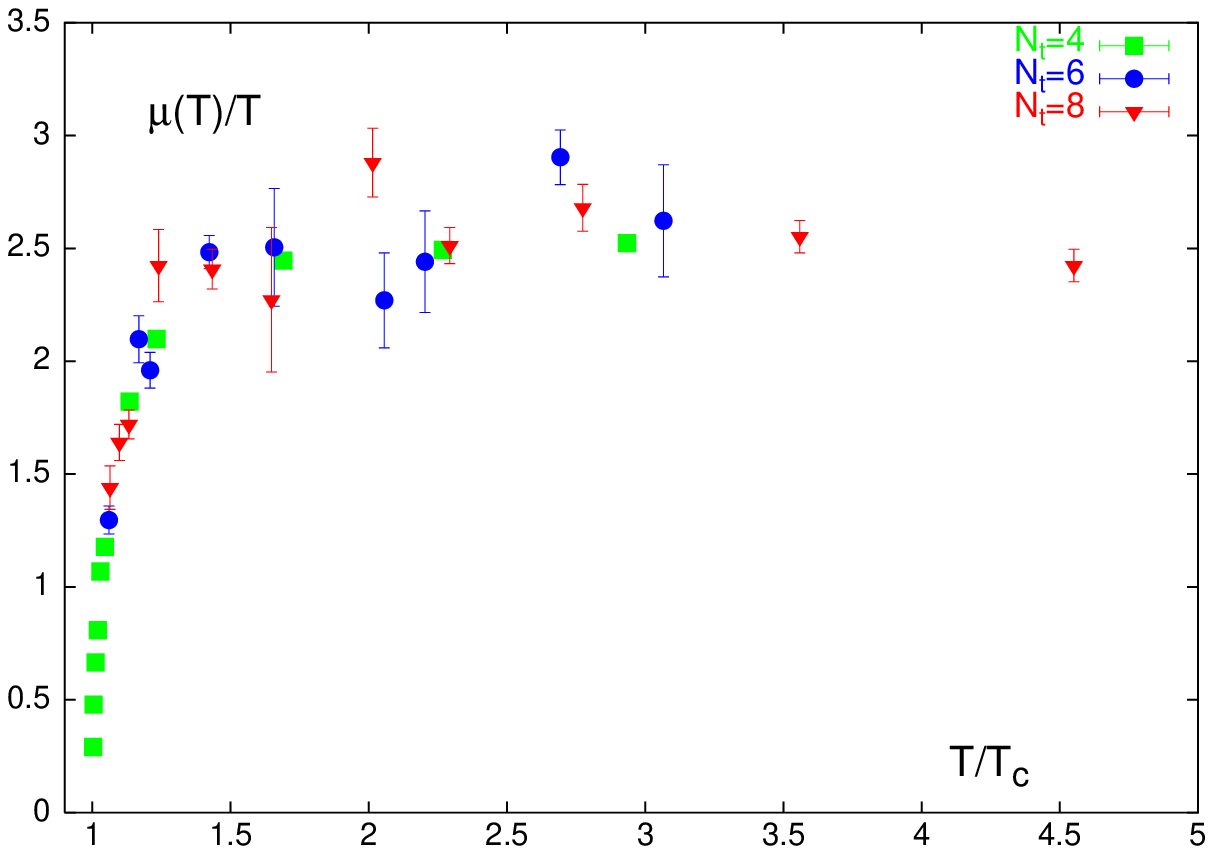,height=46mm}
\caption{Left: Static quark free energy for $N_f=3$
at temperatures $0.58<T/T_c<1.15$ \protect\cite{Peikert}. 
$F(r)$ is
normalized at ($r=1/T$) to the $T=0$
Cornell potential, $V(r)/\sqrt{\sigma} = -\alpha /
r\sqrt{\sigma} + r\sqrt{\sigma}$ with $\alpha = 0.25\pm 0.05$ (solid band).
Right: The (quenched) screening mass for $T>T_c$ from fits 
to \eq (\ref{eq:fit}) with $d=3/2$ \cite{Okacz}.
}
\label{fig:nf3potential}
\end{figure}

In the pure gluon theory, at temperatures
below $T_c$ 
and large distances $R$, the free energy rises 
linearly. The
coefficient of the linear term, the string tension,
decreases with increasing temperature and vanishes above
$T_c$ \cite{Okacz}.
In the presence of dynamical quarks the color charges
of the heavy quarks are screened also below $T_c$ and one
observes \cite{DeTar_pot,Peikert,Bornyakov}
the expected string breaking,
Figure~\ref{fig:nf3potential}.
The distances where the free energies become flat in $R$
range from 1.5 to 1 fm, decreasing with temperature,
even at (bare) quark masses as large as $m_q/T = 0.4$.
Note that the deviations from the zero temperature quenched
potential set in already at distances of
${\mathcal O}(0.5 {\rm fm})$ for temperatures $\gsim \, 0.75 T_c$.
Normalizing the free energy to the short distance
zero temperature potential, its large $R$ asymptotic value
rapidly decreases with $T$.

In the deconfined phase the free energy shows the
behavior of an exponentially screened potential, 
and is well described by
\be \label{eq:fit}
\frac{F_{Q \bar Q} (R,T)}{T} = -\frac{c(T)}{(RT)^d}\ex^{-\mu(T) R},
\ee
where $c(T),d,\mu(T)$ are fit parameters.
In perturbation theory, the leading term originates
from two-gluon exchange and predicts 
$d=2$ and exponential decay with
twice the Debye mass \cite{Nadkarni}. 
Lattice investigations \cite{Okacz,hkr}
have shown that this simple behavior does not apply in the temperature range explored.
Rather, 
fits \cite{Okacz} to \eq (\ref{eq:fit}) favored $d \simeq 3/2$ and
found screening masses $\mu(T)$ 
to be compatible
with the lowest color singlet $0^{++}_+$ screening mass
shown in Figures \ref{screen}, \ref{drcomp}. This is not surprising: 
the Polyakov loop
is a gauge invariant operator, and since it is an exponential of gauge fields
it couples to all $J^{PC}$ sectors. Consequently its correlator decays with the lightest
screening mass of the spectrum.

In order to address Debye screening of a static 
quark-antiquark system, one needs to access the singlet free energy channel, which 
in the deconfined phase is expected to decay exponentially with the electric mass $m_E$
\cite{Nadkarni1}. Constructing thermal correlators for the different channels separately
in general requires gauge fixing. Recently it has been shown that 
this can be achieved in a manner respecting the transfer matrix, and such correlators
decay with a gauge invariant spectrum \cite{op3}.
The octet channel was found to be repulsive at short distances,
as predicted by perturbation theory \cite{Brown, Nadkarni}, while at large distances
it approaches the confining singlet potential from above. 
First studies of 
the thermal singlet channel have been reported in \cite{new_okacz}.

\section{Susceptibilities}
{\label{sec:susc}

Susceptibilities of quark number and other quantities offer an elegant and simple way to 
study effects of small finite baryon densities on the equation of state
within simulations performed at zero chemical
potential \cite{oldchi}. 
Moreover, some of them are directly related to event-to-event fluctuations of particle 
production in heavy ion collisions, and help to providing signals to differentiate between
the phases in order to detect the quark gluon plasma \cite{ev}.

Quark number densities and susceptibilities are defined by
\be
   n_f(T,\mu_u,\mu_d,\mu_s)=\frac{T}{V}\frac{\partial \ln Z}{\partial \mu_f},\qquad
   \chi_{ff'}(T,\mu_u,\mu_d,\mu_s) = \frac{T}{V}
        \left.\frac{\partial^2 \ln Z}{\partial\mu_f\partial\mu_{f'}}\right |_{\mu=0},
\label{susc}\ee
where the $\mu_f$ are chemical potentials for quark number of flavor $f$, and $n_f=0$ for
$\mu_f=0$. 
Susceptibilities for conserved charges are derived by writing down the appropriate
superposition of quark chemical potentials and applying the chain rule.
First we discuss the flavor singlet and triplet quark number susceptibilities in the two flavor
case,
$\chi_{sing}=2\chi_{uu}+2\chi_{ud}, \chi_{trip}=2\chi_{uu}-2\chi_{ud}$.
Numerical studies with dynamical staggered quarks show
a rapid jump at the critical temperature \cite{got97,HT_ua1}. 
A more recent study \cite{milc} for 2+1 flavors is shown in Figure \ref{mresp},
where the strange quark mass was fixed by tuning $m_{\eta_{ss}}/m_\phi$ to its physical value,
i.e.~$m_s^{\overline{MS}}(2 {\rm GeV})\approx 95$ MeV \cite{Wittig}.
In addition it contains the diagonal strange quark number susceptibility,
which at large temperatures should become equal to 
half the triplet susceptibility.
The use of improved actions (one loop Symanzik for gauge fields and Asqtad for fermions)
leads to good apparent scaling behavior, 
suggesting deviations from the free quark gas value at $2 T_c$
of not more than 10 \% in the continuum limit.
\begin{figure}
\leavevmode
\epsfysize=5.7cm
\epsfbox{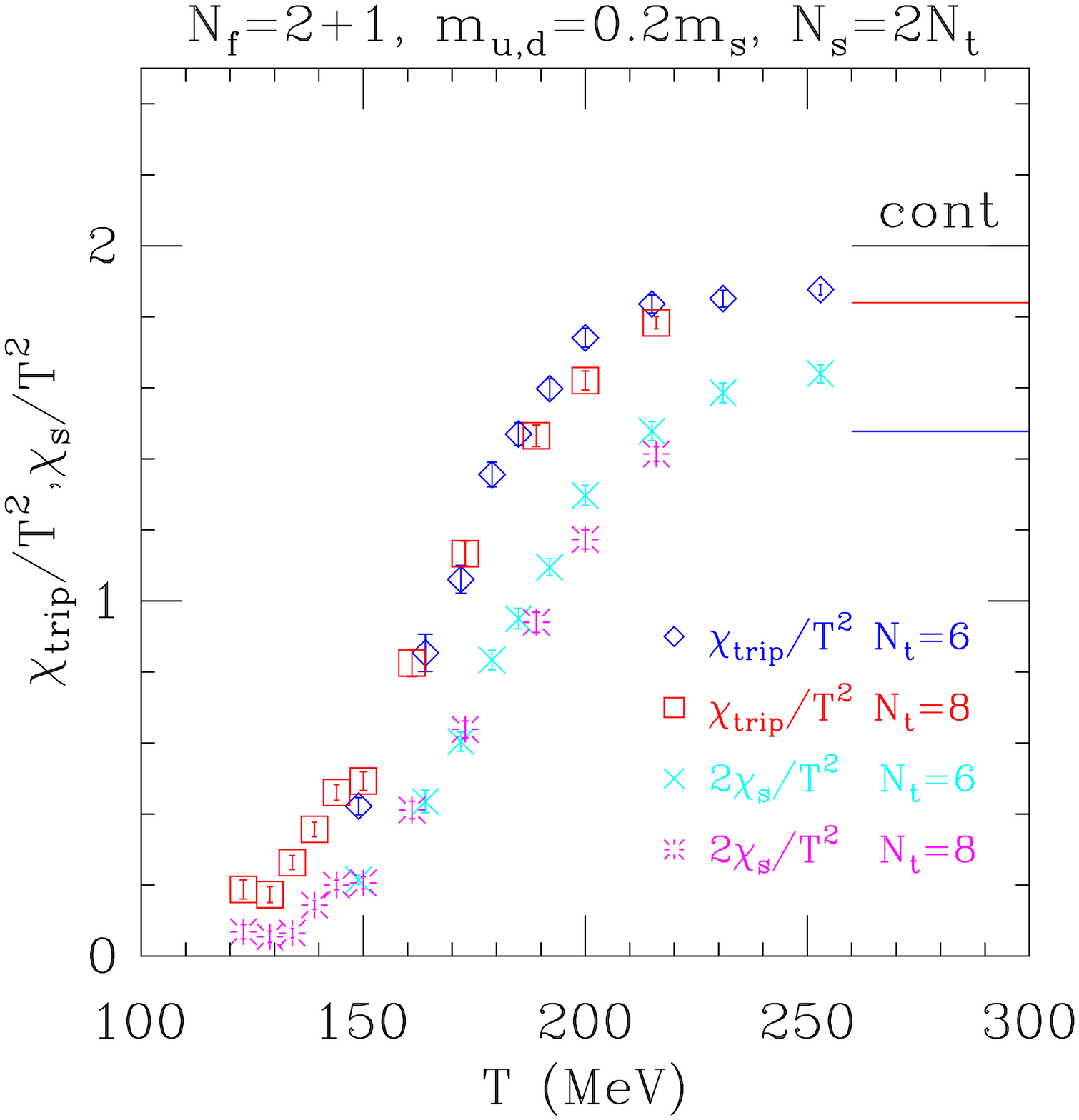}
\hspace*{1cm}
\epsfysize=5.5cm
\epsfbox{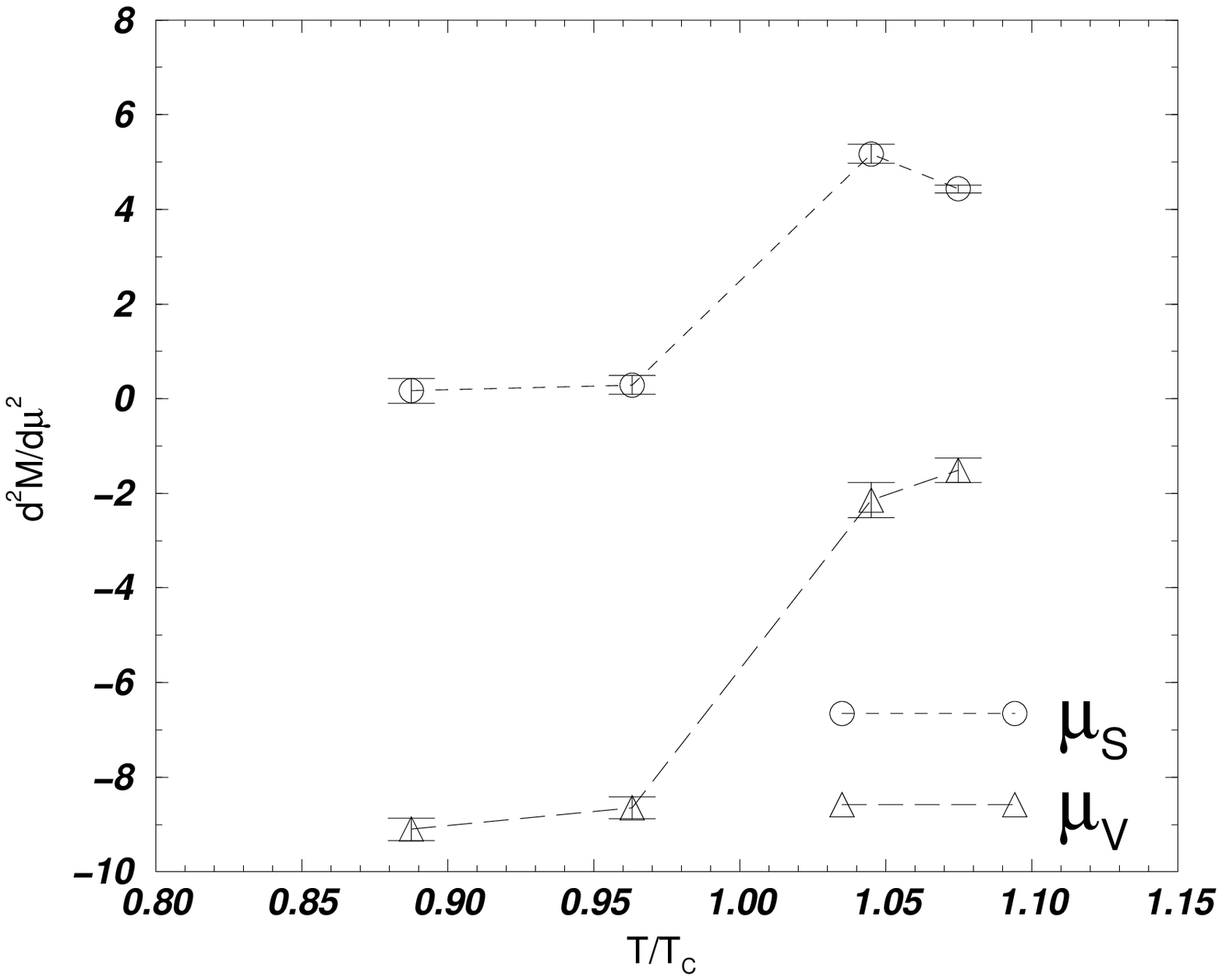}
\caption[]{Left: Triplet and strange quark number susceptibilites on $12^3\times 6$ and
$16^3\times 8$. The lines on the right indicate free quark values for $N_t=6,8$ and the
continuum \cite{milc}.
Right: Response of the pseudoscalar screening mass to chemical potentials,
$16\times 8 ^2\times 4$, $N_f=2$, $am=0.025$ ($m_\pi\approx 300$ MeV)~\cite{taro1}.}
\label{mresp}
\end{figure}

Other studies \cite{ggm,Sourendu2} considered mostly the deconfined phase, $T\geq 1.5T_c$. 
They find that in this regime quenching only affects 
susceptibilities by about 5\%, while cut-off effects for unimproved actions
play an important role. 
Based on these observations, quenched calculations using two different 
fermion discretizations were used to obtain continuum extrapolated results \cite{Sourendu2}, 
as shown in Table \ref{tb.res}. These authors 
also
study susceptibilites for baryon number,
$\chi_B = \left(4\chi_3+\chi_s+4\chi_{ud}+4\chi_{us}\right)$, 
and electric charge, 
$\chi_Q = \left(10\chi_3+\chi_s+\chi_{ud}-2\chi_{us}\right)$, 
where $\chi_3$ 
is $\chi_{trip}/2$.
Flavor off-diagonal susceptibilities
are found to be consistent with zero, which is in contrast with predictions from a
three-loop perturbative calculation \cite{bir}. It is also inconsistent with the results
of \cite{milc}, which reports statistically significant off-diagonal elements of the order
of $10^{-2}$ at $T\sim 1.5 T_c$. While this 
issue needs clarification, off-diagonal values in any case seem to be small
compared to the diagonal ones except very close to $T_c$ \cite{milc}.

For the isovector susceptibility $\chi_3$ numerical results
are compatible.
At $T=3T_c$, $\chi_3$ from Table \ref{tb.res} 
falls short of its free fermion gas value $\chi^3_{FF}=T^2$ by about 10\%.
For such temperatures and larger, it is consistent with
HTL-resummations, which predict $\chi_3/\chi^3_{FF}\approx 0.90-0.94$. This observation 
fits to the behavior of the pressure and the discussion of screening masses 
at comparable temperatures, 
with soft gluon modes accounting for remnant interactions.
It is interesting to note that for the
isovector chemical potential $\mu_3=\mu_u-\mu_d$
the fermion determinant is positive,
thus it can be simulated (cf.~Section \ref{sec:mu}) 
and numerical results can be compared to analytic 
predictions \cite{ss}.

\begin{table}
\caption{Results for the continuum limit of quark number susceptibilities
   in quenched QCD, extrapolated from staggered quark simulations 
   with valence mass $m_s/T_c=1$. From \cite{Sourendu2}.
   }
\begin{center}\begin{tabular}{@{}cccccc@{}}
   \hline\hline
   $T/T_c$ & $\qquad\chi_3/T^2$ &
         $\qquad\chi_{ud}/T^2$ & $\qquad\chi_s/T^2$ &
         $\qquad\chi_0/T^2$ & $\qquad\chi_Q/T^2$ \\
   \hline
   1.5 & 0.84 (2) & $\quad(-2\pm3)\times10^{-5}$ & 0.53 (1) & 0.43 (1)
                  & 0.99 (2) \\
   2.0 & 0.89 (2) & $\quad(-4\pm4)\times10^{-6}$ & 0.71 (2) & 0.47 (1)
                  & 1.07 (2) \\
   3.0 & 0.90 (3) & $\quad(2\pm2)\times10^{-6}$  & 0.84 (3) & 0.49 (1)
                  & 1.09 (3) \\
   \hline
\end{tabular}\end{center}
\label{tb.res}\end{table}

In a similar vein, responses of the lowest pseudoscalar screening mass to scalar and
vector chemical potentials have been investigated in a dynamical simulation with two light
staggered flavors \cite{taro1}. 
Because of the symmetry of the grand canonical partition function under combined
euclidean time and $\mu$-reflections, terms
odd in $(\mu/T)$ vanish for screening masses, 
which is also observed numerically.
The result for the leading quadratic coefficient is shown in Figure \ref{mresp}, where 
$\mu_S=\mu_u=\mu_d$, and $\mu_V=\mu_u=-\mu_d$.
Again one observes a pronounced jump across the deconfinement transition. Below $T_c$,
the response to a small scalar chemical potential is essentially zero, consistent with the
fact that pions are still Goldstone bosons associated with chiral symmetry breaking
in this regime. This ceases to be the case above
$T_c$, where a large response is observed. The behavior with isovector chemical potential
is quite different, and consistent with expectations based on the phase structure conjectured
in \cite{ss}.

\section{Dynamic Properties}

Lattice simulations inevitably are carried out in euclidean
space-time. In order to make predictions for real-time processes, 
in general an analytic continuation to Minkowski space
has to be performed. At zero temperature, 
this is trivial for
certain quantities such as masses 
of stable states
or certain matrix elements.
At finite temperature the situation is 
more complicated \cite{bellac}, since 
e.g.~the momentum space propagator is defined only
at discrete Matsubara frequencies $i \omega_n$.
Furthermore,
because Lorentz invariance is lost due to the presence
of the heat bath, spatial and temporal correlation functions 
are different in general,
and the static results
as discussed in Section~\ref{sec:screen}
can not immediately be used 
in the interpretation of experimental results 
for dynamical processes.

The full information about plasma excitations, the presence of
genuine particle poles, resonances, their location and widths etc.,
is contained in the spectral density $\sigma_H(p_0, \vec p)$
in a given channel with quantum numbers $H$.
It is related to temporal correlation functions by
\begin{equation}
G^T_H(\tau, \vec p) =
\int_{0}^{+\infty} \frac{d p_0}{2 \pi} 
\sigma_H(p_0, \vec p) 
\frac{\cosh[p_0(\tau-1/2T)]}{\sinh(p_0/2T)}\;.
\label{eq:temp_corr}
\end{equation}

Temporal correlations have been used in order to gain
information about the spectral density.
However, since the extension of the system in the
temporal direction is limited by the inverse temperature,
the isolation of a ground state dominating the
correlation function at large distances is difficult.
To improve the reliability of fits,
extended 
(smeared) 
operators of various kinds have been used, sometimes combined
with anisotropic lattices
\cite{Nucu_400,Nucu_1}.
With smearing one hopes to increase the projection onto 
the ground state, whose contribution would then dominate
already at small temporal
distances. Anisotropic lattices
increase the number of data points in the temporal direction, to
stabilize more sophisticated fit ans\"atze
including ground and first excited states.
However, one still has to rely on fit ans\"atze
and, if possible, use the quality of the fit to distinguish between
various models.
Moreover, modeling the operators introduces bias, and
the results have thus turned out to depend on the method.
However, at least qualitatively the results
are in agreement with 
a significant two-quark cut contribution,
while in addition the behavior of wave functions
was interpreted to indicate meta-stable bound states in the plasma
phase \cite{Nucu_1}.

Clearly, it would be much more preferable to obtain 
the spectral density $\sigma_H$ directly.
However, since the lattice provides only a limited
set of noisy data points at discrete values of $\tau$, its 
numerical extraction from correlation functions is
an ill-posed problem.
Here, progress has been made recently \cite{Asa00}
by applying the
maximum entropy method (MEM) \cite{Bryan}, which attempts
to construct the most probable spectral density given the
data and taking into account prior knowledge, such as the
positivity of $\sigma_H$ and its perturbative behavior 
at large energies.
The method has been successfully tested and applied 
at zero temperature \cite{Asa00,CPpacs02},
where the contributions of several excited states
to the spectral density could be established.
At finite temperature one still is hampered by
the limited temporal extent of the lattice.
Moreover, it is important to check whether one is able
to reproduce the two-quark cut in the free case
\cite{Wet00}, because this is expected to contribute
significantly also in the interacting
case at high temperature.
First results for the vector channel at small quark 
masses \cite{Kar02a} 
are shown in Figure \ref{fig:vector}.
The main feature of the data is that the rho peaks
at low temperature disappear and develop into broad ``resonances''
above $T_c$ whose locations move proportional
to the temperature.
The vector spectral density $\sigma_V(p_0, 0)$ is
immediately related to the thermal cross section for the 
production of dilepton pairs at vanishing momentum \cite{Braaten},
\begin{equation}
{{\rm d} W \over {\rm d}p_0 {\rm d}^3p}|_{\vec{p}=0} =
{5 \alpha^2 \over 27 \pi^2} {1\over p_0^2 ({\rm e}^{p_0/T} - 1)}
\sigma_V(p_0,\vec{0})\;.
\end{equation}  
This is shown in Figure \ref{fig:vector} (right).
The ``resonance'' like enhancement of $\sigma_V$ results
in the enhancement of the dilepton rate over the perturbative tree 
level (Born) rate \cite{Kapusta_dilep} for energies $p_0 / T \in [4,8]$.
In the low energy range the spectral density decreases rapidly,
in contrast to perturbative calculations
which have $\sigma_V$ diverging in this limit \cite{Braaten}.
Fortunately,
from inspecting the correlation function
directly \cite{Kar02a} it is already clear that the spectral density vanishes
$ \sim p_0^\alpha$ with some power $\alpha$ in the $p_0 \rightarrow 0$
limit. 
However, gaining detailed control over the low energy
behavior of the spectral density is a challenge at this point
and will become even more difficult at higher temperatures.
On the other hand, the zero energy limit of spectral functions
is related to transport coefficients \cite{Aarts},
and it would be extremely interesting to obtain results
for these quantities \cite{Sou_trans} as a step into 
first principles investigations
of non-equilibrium properties of QCD.

\begin{figure}
\begin{center}
\epsfig{file=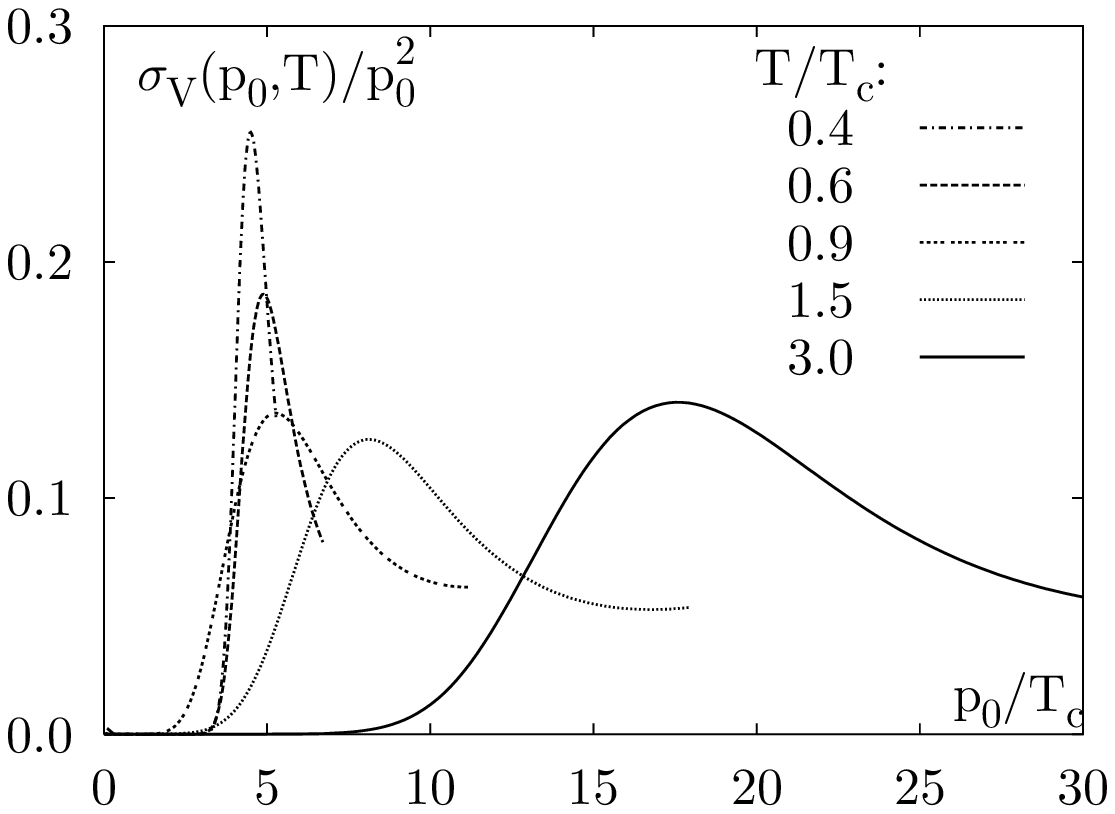,width=65mm}
\epsfig{file=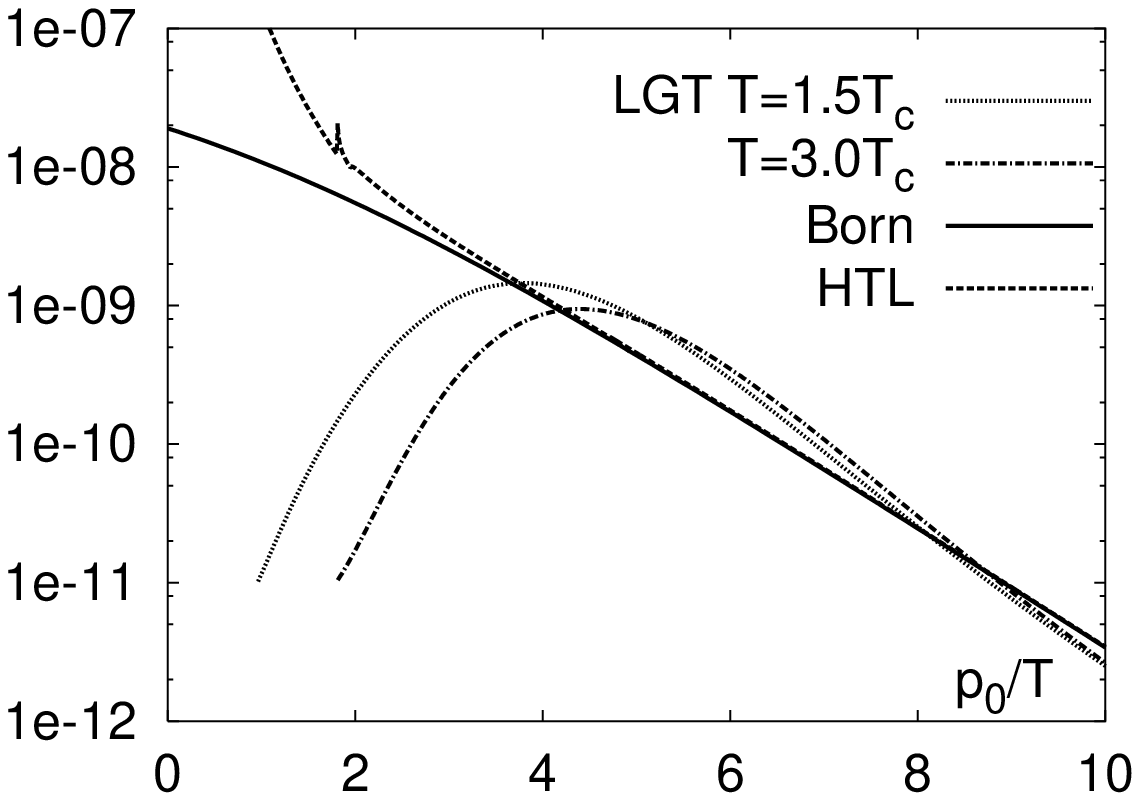,width=65mm}
\end{center}
\caption{Left: Vector spectral functions for various $T$.
Right: Dilepton rate 
at $1.5$ and $3 T_c$, together
with tree level and hard thermal loop prediction.
From \cite{Kar02a}.
}
\label{fig:vector}
\end{figure}

\section{Lattice QCD at Finite Density}
\label{sec:mu}

QCD at finite baryon density plays a role in two rather different regimes.
In heavy ion collisions the initial state has a small non-zero baryon number,
corresponding
to baryon chemical potentials of order 
$\sim \op(50\,{\rm MeV})$ \cite{exp}, 
and any subsequent plasma state is a state of high temperature and low density.
On other hand, the core of neutron stars is composed of cold and very dense nuclear matter. 
These two situations correspond to the regions close to the axes in the
tentative QCD phase diagram Figure~\ref{pdiag}.
\begin{figure}[htb]
\vspace{9pt}
\begin{center}
\includegraphics[width=7cm]{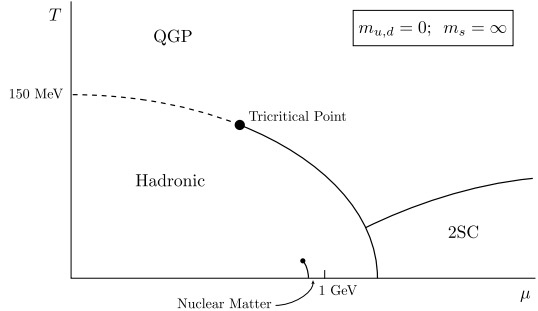}
\end{center}
\vspace*{-0.5cm}
\caption[]{\label{pdiag}
Conjectured phase diagram for QCD with two massless and one heavy flavour. From \cite{kr}.}
\end{figure}
While in the latter case a rich phase structure including a
color superconducting state has been conjectured \cite{csc}, understanding the former regime
is particularly pressing in view of current heavy ion collison experiments.
Of special importance is a prediction of the   
location of the tricritical point, in which the first order deconfinement 
transition line is expected to terminate.


Simulations of lattice QCD have so far failed to be a viable tool for
finite density because of the so-called ``sign problem''. 
For $\mu=0$ the relation
$\gamma_5 M\gamma_5=M^\dag$
guarantees positivity of the fermion determinant, $\det M(0) \geq 0$.
However, for SU(3) and $\mu\neq 0$ this relation does not hold and 
$\det M$ is complex. This prohibits Monte Carlo importance sampling, 
which interprets the measure as a probability factor and thus requires it to
be positive.  
Note that all physical quantities like $Z$ or observables $\langle \op\rangle$ are real 
positive, i.e.~all imaginary parts cancel out of the path integral 
and the problem lies merely in evaluating it.

This problem is generic for many fermionic models, and for a certain class of
them it can actually be solved by the use of 
cluster algorithms. These subdivide the ensemble into parts within which all
contributions add up positively. The model closest to QCD for which this succeeded
is the three state Potts model \cite{pott}.
Unfortunately, these algorithms are not directly applicable to QCD, and we 
refer to \cite{clus} for an overview.

We will likewise not discuss the considerable amount of work done on certain modifications
of QCD for which the sign problem is manageable or altogether absent, such as
QCD in the static quark limit, 
two-color QCD or QCD at finite isospin. These topics are covered in recent reviews
\cite{latrevs}. 
It has also been argued that for very high densities an effective
theory free of the sign problem can be derived \cite{hsu}, but no numerical results
exist as of yet.

Here we wish to focus on some recent approaches 
which made significant progress towards simulating
QCD with realistic parameter values at small baryon densities. It is important to note
that all of them circumvent the sign problem rather than solving it.
The partition function in its first form in \eq (\ref{lagrangian}) suggests 
that for small $\bmu=\mu/T$ the problem is close to the trivial case, 
and in particular lends itself to a Taylor expansion. The following approaches 
in one way or another all require small enough $\bmu$ to succeed. 
As we shall see, however, this limitation still allows to obtain results up to
$\mu_B\sim 500$ MeV, thus fully including 
the parameter region relevant for heavy ion collisions. In the following we
limit our discussion to mapping out the deconfinement transition line. 

\subsection{Reweighting: The Glasgow Method}

A mathematical identity can be used to reweight 
from configurations generated at zero density, where there is no sign problem,
to a corresponding one at finite density. The method is based on rewriting the partition
function as
\be \label{rew1}
Z=\int DU\, \det M(0)\, \frac{\det M(\mu)}{\det M(0)}\,\ex^{-S_g[U]}
=\left\langle\frac{\det(M(\mu))}{\det(M(0))}\right\rangle_{\mu=0},
\ee
where the determinant ratio is now treated like an observable while the integration measure is
defined at $\mu=0$, and hence positive.
This expression is exact. However, in a Monte Carlo integration with finite
statistics two problems arise. Firstly, the complexity of the numerator leads to 
oscillations and large cancellations.
The reweighting factor corresponds to a ratio of two partition functions with
different actions, and thus decays exponentially with the
difference between their free energies,
which is an extensive quantity, $Z=\exp{-\Delta F/T}\sim \exp{(-const.V)}$.
The statistics required for a given accuracy thus grows exponentially
with volume, rendering extrapolations to the thermodynamic 
limit extremely difficult.
Secondly, Monte Carlo approximates the partition function 
by the contributions of the most 
likely configurations obtained by probability sampling. Since the extrema of the action 
with $\mu=0$ are shifted from those of the entire integrand
including the determinant ratio, the probability distribution of the generated ensemble is
changed. This should be of little consequence for small $\mu$ and large statistics,
but with growing $\mu$ the overlap between reweighted and full ensemble deteriorates.
The effect does not show up in the statistical error, 
and the difficutly is in knowing when this becomes problematic leading 
to incorrect results. For example, at $T=0$ the transition to nuclear matter
is found at an unphysical value for $\mu_c$.
A detailed discussion with early results and references can be found
in \cite{bar}.

\subsection{Multiparameter Reweighting}

Only recently it was realized that the ensemble overlap can be significantly improved by 
a multidimensional generalization \cite{fk1} of the Glasgow method.
In addition to $\mu$ one may also reweight in the lattice
gauge coupling $\beta$, so that the partition function is now written as
\be
Z=\left\langle \frac{\ex^{-S_g(\beta)}\det(M(\mu))}
{\ex^{-S_g(\beta_0)}\det(M(0))}\right\rangle_{\mu=0,\beta_0}.
\ee
This is crucial for finding critical behavior.
While reweighting only in $\mu$ would inevitably
generate an ensemble away from criticality, the second reweighting parameter
can be used to keep it fluctuating between the phases, as the
unshifted ensemble certainly would. 
This method was compared with the Glasgow method
in four flavor QCD at imaginary chemical potential \cite{fk1}, 
for which the determinant is positive and a full
ensemble is also available (cf.~Section~\ref{imag}). 
Multiparameter reweighting is found
to agree significantly better with the full ensemble up to some 
critical imaginary chemical potential,
beyond which it breaks down \cite{fp}.

The method was then used in \cite{fk2} to map out the deconfinement line, 
using Lee Yang zeros \cite{ly} as observables. 
These are points of vanishing partition
function, marking the singularities in the free energy. Singular phase transitions only 
appear in the thermodynamic limit, while on finite volumes the free energy is regular.
In the latter case the Lee Yang zeroes are displaced to complex parameter values.
As the volume is increased, they approach their thermodynamic limit values on the real axis
if there is a genuine phase transition, whereas for a smooth crossover the zeroes settle
somewhere in the complex plane. In the case of a first order transition the large volume 
behavior is consistent with a $\beta_c(V)=\beta_c(\infty)+\zeta/V$ behavior.
The simulations reported in \cite{fk2} were carried out for 
2+1 flavors of staggered fermions, the result is shown in Figure \ref{comp}.  
It constitutes the 
first numerical prediction of the $(\mu,T)$ 
phase diagram including a critical point.  
However, larger volumes are required to check for thermodynamical behavior
and confirm the results. Recently attempts have been made to further
improve the overlap by splitting the reweighting factor in a product
of many factors, and preliminary results 
on small lattices have been reported in \cite{crompton}.

\subsection{Taylor Expanded Reweighting}

Since for heavy ion collisions one is interested in rather small 
chemical potentials
of a few ten MeV, one may use the smallness of $\bmu$ 
in an approximation of multiparameter
reweighting in order to make simulations
on larger volumes feasible \cite{allt}. Consider a Taylor expansion of 
the reweighting factor ${\cal R}$ as a power series in $\bmu=\mu/T$,
and similarly for any operator $\op$. Expectation values are then given by
\be
\langle{\cal O}\rangle_{(\beta,\mu)}=
{
{\langle({\cal O}_0+
 {\cal O}_1\bmu+{\cal O}_2\bmu^2+\ldots)
  \exp({\cal R}_1\bmu+{\cal R}_2\bmu^2+\ldots-\Delta S_g)\rangle_{\bmu=0,\beta_0}}
\over
{\langle\exp({\cal R}_1\bmu+{\cal R}_2\bmu^2+\ldots-\Delta S_g)
\rangle_{\bmu=0,\beta_0}}}.
\ee 
For sufficiently small chemical potentials, this series should converge quickly and
calculation of the leading coefficients should give a good approximation to the full answer.
The benefit is that the derivatives of the determinant can be expressed through local operators 
in the fermion fields. These are much cheaper to compute than the whole 
determinant, 
which requires a non-local operation. In particular, keeping only the leading term, the cost is
equivalent to that of computing susceptibilities (cf.~Section \ref{sec:susc}). 
Even though the expanded
phase of the determinant still has oscillations growing with volume and $\bmu$,
the saving in computer time can be used to gather higher statistics for larger volumes.

This strategy has been applied in \cite{allt} to compute the location of 
the deconfinement line for a two-flavor theory with 
p4-improved staggered fermions and 
a pion mass of about $m_{\pi}\approx 600$ MeV. 
The method to locate a phase transition is to look for the peak
$\chi_{max}=\chi(\mu_c,\beta_c)$
of the susceptibilities $\chi= VN_t \left\langle(\op - \langle\op\rangle)^2\right\rangle$,
where the chiral condensate and the Polyakov loop were
used for $\op$ in practice.
This peak implicitly defines a critical coupling $\beta_c(\bmu)$, which must be even in $\mu$
because of the symmetry of the partition function \cite{fp}. 
This can be employed to calculate the curvature of
the critical temperature at $\mu=0$,
\be
\frac{d^2T_c}{d\mu^2}=-\frac{1}{N^2_tT_c}\frac{d^2\beta_c}{d\mu^2}
\left(a\frac{d\beta}{da}\right)^{-1}.
\ee

Indeed a lattice with twice the
spatial size than that in \cite{fk2} was employed without major problems.
Of course, in order to judge the convergence properties of the series, several 
consecutive coefficients are needed.

\subsection{\label{imag}Imaginary Chemical Potential}

For imaginary chemical potential, the fermion determinant is positive and
simulations are as straightforward as for $\mu=0$. It is therefore natural
to ask whether such simulations can be exploited to learn something about real
$\mu$.  Let $\mu_R,\mu_I\in {\rm I\kern -.2em  R}$ denote the
real and imaginary parts of a complex $\mu = \mu_R + \ii \mu_I$. 
In the presence of a complex chemical potential,
a $Z(3)$ transformation of
the fermion fields is equivalent to a shift in $\mu_I$, causing the partition function 
to be periodic in $\mu_I$ with period $2\pi T/3$ \cite{rw},
\be \label{s2}
Z(\mu_R,\mu_I)=Z(\mu_R,\mu_I+2\pi T/3).
\ee
Hence, once $\bmu_I$ exceeds some critical value $\bmu_I^c$, a phase
transition to a non-trivial $Z(3)$ sector occurs, and because of the symmetry \eq (\ref{s2})
this transition is periodically repeated at the exact critical values
$\bmu_I^c=2\pi (k+1/2)/3$. 
It was conjectured \cite{rw} and verified numerically \cite{fp,el} that these transitions
are of first order in the deconfined phase and continuous in the confined phase.
On the other hand, for purely real $\mu$ these transitions are absent.

Through its periodicity, the grand canonical partition function at imaginary
chemical potential is the Fourier transform of the
canonical partition function at fixed quark number $Q$ \cite{rw,immu},
\be
Z_Q(T,V)=\frac{1}{2\pi}\int_0^{2\pi}d\mu_I \,\ex^{-\ii\mu_I Q/T}Z(\mu=\ii\mu_I,T,V).
\ee
It was thus suggested to simulate $Z$ at imaginary chemical potential and
perform the Fourier transformation numerically \cite{akw}. This has been
carried out in the 2d Hubbard model at large $T$ and small $Q$, but not yet 
in QCD. Unfortunately, this approach only postpones the sign problem from the Monte
Carlo integration to the Fourier transform. As $Q$ is getting large approaching the
thermodynamic limit 
at fixed baryon density, it will again cause rapid oscillations and the 
associated cancellations.

\subsection{Analytic Continuation}

If $\bmu$ is small another line of attack is possible \cite{ml}.
Since there are no massless modes in the theroy, all observables are analytic
functions of $\bmu$ everywhere but on the critical lines of phase transitions.
After computing expectation values with $Z(\ii\mu_I,T,V)$, one may therefore
fit them by truncated Taylor series 
\be
\langle \op \rangle = \sum_n^N c_n \bmu_I^{2n}.
\ee
In cases where this is possible to satisfactory accuracy, 
analytic continuation to real $\mu$ is trivial.
There are two advantages to such a procedure.
First, there is no sign problem and hence
no restriction on the volumes that can be simulated. Second, the simulation gives
the value of the whole expectation value rather than just a single Taylor coefficient.
While the expansion is necessary in order to analytically continue, fitting to various
orders allows control over the convergence properties.
However, a check of the convergence and hence controlled continuation are limited to 
$\bmu_I<\bmu_I^c=\pi/3$, which marks 
the first $Z(3)$ transition discussed above. 
 
Non-perturbative evidence for the viability of this approach has been given for
screening masses in the deconfined phase \cite{hlp2}, which can be simulated
sucessfully for imaginary and real $\mu$ in
the framework of dimensionally reduced QCD \cite{hlp1}.

It was shown in \cite{fp} that the same approach can be extended to study the 
critical line of the deconfinement transition itself.
Again the transition was defined by the peak
$\chi_{max}=\chi(\mu_c,\beta_c)$
of plaquette, chiral condensate or Polyakov loop susceptibilities. 
On finite volumes these are analytic
functions over the whole parameter space of the theory,
non-analyticities associated with phase transitions develop only in the thermodynamic
limit.
The implicit function theorem then implies that also
$\beta_c(\bmu)$ is analytic for all $\bmu$, which therefore has a Taylor expansion
\be \label{beta}
\beta_c(\bmu)=\sum_{n}c_n \bmu^{2n}.
\ee

As the thermodynamic limit is approached, 
$\beta_c(\bmu)$  approaches the unique critical line of phase transitions,
while remaining pseudo-critical in a crossover regime.
In principle the nature of the line can also be determined from its
finite volume scaling.

In \cite{fp} a two staggered flavor QCD 
was simulated, and a striking finding
is that for all $\bmu_I<\bmu_I^c=\pi/3$ the critical coupling is indeed well described
by a fit quadratic in $\bmu_I$, with negligible effect of terms ${\cal O}(\bmu_I^4)$.
The same conclusion is reported from an imaginary $\mu$ simulation of a four flavor theory
in \cite{el}. In principle, of course, there is nothing that guarantees such good behavior,
which would need to be checked for every change of parameters.
It is therefore amusing to note that both in the 3d Gross-Neveu model
as well as in some random matrix model with symmetries similar to QCD, the leading
quadratic term very well reproduces the exact soultion for the critical line
over wide ranges of parameter space \cite{el}.

\subsection{The $(\mu,T)$ Phase Diagram}

In Figure \ref{comp} we assemble the various phase diagrams published so far in one plot.
Some caution shoud be used when interpreting it, as the curves correspond to different
parameter values and volumes. 
The only thing equal in all four simulations is the lattice spacing
which at $a\approx 0.3$ fm is large enough to cause cut-off effects.
Furthermore, the error bars shown on the
data points should not be directly compared with the 
smaller error bands corresponding to fit results.  
\begin{figure}[th]%
\vspace*{0.5cm}
\begin{center}
\includegraphics[width=7cm]{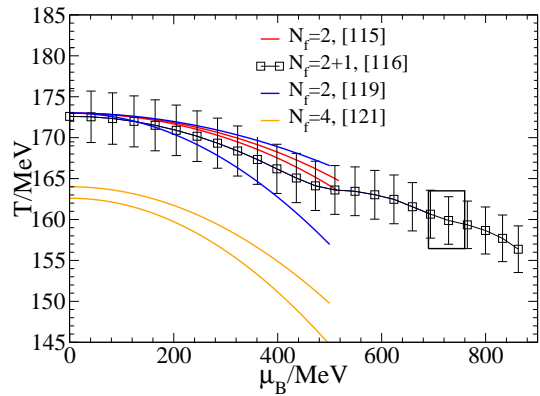}
\end{center}
\caption[a]{\label{comp}
Location of the (pseudo-)critical line as reported in four different 
simulations. All employ staggered fermions, 
only Allton et al.~use improved actions. 
The simulation parameters are collected in Table \ref{params}.
}
\end{figure}
\begin{table}
\caption[]{Simulation parameters for the results shown in Figure \ref{comp}.}
\begin{center}
\begin{tabular}{@{}cccc@{}}
\hline \hline
Method & $N_f$ & $m_q$ & largest lattice \\
\hline
reweighting \cite{fk2} & 2+1 & $am=0.025, m_s=8m_u$, $(m_{\pi}\approx 300 MeV)$ & $8^3\times 4$\\
rew.~+Taylor \cite{allt} & 2 & $am=0.1$, $(m_{\pi}\approx 600)$ MeV & $16^3\times 4$ \\
imag.~$\mu$ \cite{fp} & 2 & $am=0.025$, $(m_{\pi}\approx 300)$ MeV & $8^3\times 4$\\
imag.~$\mu$ \cite{el} & 4 & $am=0.05$ &$16^3\times 4$\\
\hline
\end{tabular}
\end{center}
\label{params}
\end{table}

Nevertheless, the curves are only shown as far as the validity of the
quadratic approximation has been checked explicitly in the imaginary $\mu$ approach. 
Within this range it was found in \cite{allt} that
doubling the quark mass did not affect the line 
within the current errors, so that comparing different quark mass values
makes sense. Moreover, the four flavor calculation at imaginary $\mu$ in \cite{el}
is entirely consistent with the (unexpanded) reweighting calculation 
performed at exactly the same parameter values \cite{fk1}. Finally, 
within the Taylor expanded reweighting it was checked that simulating real
and imaginary chemical potential indeed produces the same leading coefficient \cite{ej}.

These cross-checks are extremely important, 
since all methods introduce some systematic
error growing with $\mu/T$. However,
it seems safe to conclude that they
all give compatible and reliable results for the pseudo-critical 
temperature at least up to $\mu_B\approx 500$ MeV. 
As expected on physical grounds,
$T_c(\mu)$ decreases faster when more light quarks are present. 
An important physics conclusion then is the flatness of the critical line, which implies
that the critical temperature relevant for RHIC is essentially the same as at
zero density. It also implies that $T_c(\mu_B\rightarrow 1 {\rm GeV})$ 
has to drop rather dramatically if our qualitative picture of the phase diagram
is at all correct. First investigations of the equation of state at finite
$\mu$ have found negligible effects at the few percent level for RHIC
densities \cite{fk3,allt}.

\subsection{The Critical Point, Quark Mass Dependence}

The only simulation that has so far produced a critical point
is the one employing reweighting on lattices of rather limited 
size \cite{fk2}. 
On the other hand,
experience at zero density indicates that rather large volumes 
are required in order
to reach the asymptotic finite volume scaling required for an 
unambiguous determination
of the order of a phase transition \cite{ChSchmidt}. 
Such systematic effects may well be larger
than the statistical error box indicated in the figure, 
and it is hence important
to have another independent calculation. 
Moreover, while the location of the 
(pseudo-)critical line appears to be only very weakly quark mass dependent, 
the location of the critical point strongly depends on them: 
for three chiral quarks, the first
order line joins the temperature axis, while for the 2+1 flavor case depicted in the figure
it is quite far out.

The task of locating the critical point as a function of parameters is equivalent to mapping 
out the critical surface separating first order transitions
from crossover in Figure \ref{mschem}, which is simply the extension of
the phase diagram Figure \ref{fig:phase} to finite densities. 
Probably the most economical way of doing this is to find the projections of this
surface onto the $(T,\mu)$ and various $(m_q,\mu)$-planes.
Just like the former, 
the $(m_q,\mu)$-plane then shows a critical line separating first order from
crossover. For this line, $m_q^c(\mu)$, the same analyticity and symmetry 
considerations apply as discussed for $T_c(\mu)$, i.e. it has a Taylor series even in $\mu$. 
Within the Taylor expanded reweighting approach, a first result for the leading coefficient 
in the case of three degenerate flavors of p4-improved fermions ($m_\pi \approx 200 {\rm MeV}$) 
has been reported in \cite{mc} as $T\partial m_q^c/\partial (\mu^2)=0.21(6)$.
Moreover, as argued in \cite{fp2}, when the quark mass is varied, the 3 flavor critical endpoint
will trace out a line $T_c(\mu_c,m_q)$, which has an analytic continuation to imaginary $\mu$.
This in turn allows to determine it from imaginary chemical potential simulations.
\begin{figure}[th]%
\begin{center}
\includegraphics[width=7cm]{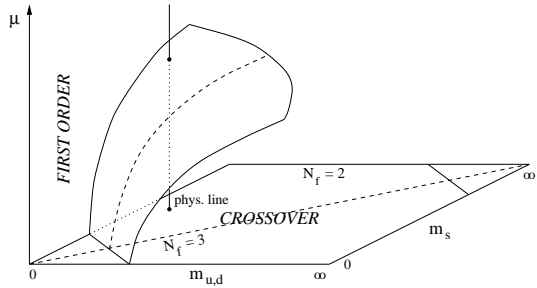}
\end{center}
\caption[a]{\label{mschem}
Critical surface separating the QCD parameter space for first order transitions from crossover.
}
\end{figure}

\section{Summary}

In this article we have attempted to summarize the current
understanding of QCD at non-vanishing temperature
as it arises from numerical studies on the lattice.
While many of the results in the pure gauge theory are
available in the continuum limit, simulations with
dynamical fermions still suffer from systematic errors.
These are mainly due to finite lattice spacing as well as quark masses
which don't meet their physical values yet.
Nevertheless, for critical temperatures and the equation
of state quantitative results could be given, which
are expected to be correct at the 10 - 15 \% level.

Apart from the basic thermodynamic quantities, existing results 
provide us with a detailed picture
of how static quantities change through the deconfinement
transition up to a few $T_c$.
Combinations of perturbative calculations and numerical
simulations have produced insight into the regime of very high 
temperatures as well as the dynamics
and mixing of modes. 
Altogether this led to a quantitative understanding of the relevant
static length scales in the plasma, as well as tests of the applicability
and breakdown of thermal perturbation theory.

The naive picture of the deconfined phase as a weakly interacting
parton gas is not supported. For temperatures relevant to
heavy ion collisions, the plasma displays strong residual interactions
through soft gluonic modes, which cannot be treated
perturbatively, and which influence different 
quantities in different ways. In particular, any constituents
themselves are objects dressed by non-perturbative
interactions. This gives a consistent explanation to the
various observed features: the equation of state, 
susceptibilities and fermionic correlators 
are dominated by hard modes, 
but significant deviations from ideal gas
behavior are still present.
Gluonic correlators, on the other hand,
are dominated by soft modes and entirely off their leading perturbative
predictions. An ideal gas is established 
only at asymptotically high temperatures.

The study of real time dynamics in the plasma is
a much more difficult problem, for which 
at present no formal approach
exists within lattice gauge theory. 
A step towards inspecting such physics could be presented by means
of maximum entropy methods. However, at this stage such calculations have
still a qualitative character, in trying to establish a working
methodology.

A large section of this paper was devoted to another new
development, allowing to study QCD also
at small baryon density, as it
is explored by heavy ion experiments at RHIC and LHC.
Again, the presented studies are still at an exploratory
level, subject to the already mentioned systematic effects.
However, rather different methods have
led to a consistent dependence of the critical temperature
on the chemical potential, giving confidence in their
validity up to $\mu_B\sim 500$ MeV. In this range the pseudo-critical
line is found to be rather flat. There also is a prediction for a critical
point to be cross checked in simulations to come.

Advances in lattice QCD have always been brought about
by a mixture of new calculational schemes, new algorithms and
refined analysis techniques. They have, however,
also been related to progress in computational
resources. With a new generation of machines in the
Teraflops-range 
going into operation soon, one has all reasons to
hope that part of the mentioned systematic deficiencies
will be reduced and simulations at parameter values
much closer to the physical ones will be realized.
We hope to have also shown that, by combining analytic and
numerical techniques, insight can be gained and new paths may show
up, which might be needed to venture fully into the real time regime.

\end{document}